\documentclass[pdflatex,sn-nature]{sn-jnl}% Style for submissions to Nature Portfolio journals
\usepackage{graphicx}%
\usepackage{multirow}%
\usepackage{amsmath,amssymb,amsfonts}%
\usepackage{amsthm}%
\usepackage{mathrsfs}%
\usepackage[title]{appendix}%
\usepackage{xcolor}%
\usepackage{textcomp}%
\usepackage{manyfoot}%
\usepackage{booktabs}%
\usepackage{algorithm}%
\usepackage{algorithmicx}%
\usepackage{algpseudocode}%
\usepackage{listings}%
\usepackage{hyperref}
\usepackage{textgreek}
\theoremstyle{thmstyleone}%
\theoremstyle{thmstyletwo}%

\theoremstyle{thmstylethree}%

\usepackage[markup=underlined]{changes}
\makeatletter
\AddToHook{cmd/added/before}{\def\Changes@AuthorColor{blue}}
\AddToHook{cmd/deleted/before}{\def\Changes@AuthorColor{brown}}
\AddToHook{cmd/replaced/before}{\def\Changes@AuthorColor{purple}}
\makeatother
\usepackage{todonotes}
\setcommentmarkup{\todo[color={pink},size=\scriptsize]{#3: #1}}

\begin{document}

\title[Article Title]{Sustained macroscopic quantum coherence in a superradiant solid under ambient conditions}


\author[1,2,3]{\fnm{Wei-Jiang} \sur{Wu}}%\email{weijiangwu@cuhk.com}
\equalcont{These authors contributed equally to this work.}

\author[1,2,3]{\fnm{Da-Wu} \sur{Xiao}}%\email{dawuxiao@cuhk.edu.hk}
\equalcont{These authors contributed equally to this work.}

\author[1,2,3]{\fnm{Wen-Tao} \sur{Wang}}%\email{1155246290@link.cuhk.edu.hk}
\equalcont{These authors contributed equally to this work.}

\author[1,2,3]{\fnm{Xin-Yu} \sur{Chen}}%\email{}
\author[1,2,3]{\fnm{Xian-Feng} \sur{Wang}}%\email{}
\author[1,2,3]{\fnm{Ming-Zhong} \sur{Ai}}%\email{}

\author[1,3,4]{\fnm{Quan} \sur{Li}}

\author*[1,2,3,4]{\fnm{Ren-Bao} \sur{Liu}}\email{rbliu@cuhk.edu.hk}

\affil[1]{\orgdiv{Department of Physics}, \orgname{ The Chinese University of Hong Kong}, \orgaddress{\street{Shatin, New Territories}, \city{Hong Kong}, \country{China}}}

\affil[2]{\orgdiv{New Cornerstone Science Laboratory}, \orgname{ The Chinese University of Hong Kong}, \orgaddress{\street{Shatin, New Territories}, \city{Hong Kong}, \country{China}}}

\affil[3]{\orgdiv{The State Key Laboratory of Quantum Information Technologies and Materials}, \orgname{ The Chinese University of Hong Kong}, \orgaddress{\street{Shatin, New Territories}, \city{Hong Kong}, \country{China}}}

%\affil[4]{\orgdiv{The Hong Kong Institute of Quantum Information Science and Technology,}, \orgname{ The Chinese University of Hong Kong}, \orgaddress{\street{Shatin, New Territories}, \city{Hong Kong}, \country{China}}}

\affil[4]{\orgdiv{Centre for Quantum Coherence}, \orgname{ The Chinese University of Hong Kong}, \orgaddress{\street{Shatin, New Territories}, \city{Hong Kong}, \country{China}}}


\abstract{Macroscopic quantum coherence, such as in laser, Bose-Einstein condensates, superfluids, and superconductors, is important to fundamental physics and useful for quantum technologies. Superradiance provides a mechanism to produce coherence among a large number of particles and photons. Its implementation, however, has been limited to gaseous systems, solids at very low temperature, or short pulses. Here we demonstrate a solid-state superradiant maser under ambient conditions, which establishes long-lived coherence among about $10^{14}$ nitrogen-vacancy center spins in diamond and about $10^9$ photons in a microwave cavity. By varying the system parameters to access the above-threshold, well-above-threshold, and deep-above-threshold regimes, we observed continuous-wave masing, periodic amplitude modulation, and sequences of superradiant bursts, which are attributed, correspondingly, to macroscopic spin coherence synchronized at a fixed frequency, a coherent time crystal of large spins, and unsynchronized superradiant transients. This work demonstrates that macroscopic quantum coherence can be spontaneously generated and maintained in solids under ambient conditions and provides a solid-state platform for exploring bright quantum lights with many-body correlations.}

\maketitle

%\section{Introduction}\label{sec1}
%\mynote{Start the story with remarks on macroscopic quantum coherence ...}
%\mynote{Introduce superradiance and superradiance laser/maser as a controllable mechanism to utilize and generate macroscopic coherence of many particles.}

Macroscopic quantum coherence arises when many microscopic constituents share a well-defined phase relation~\cite{penrose1956bose,leggett1980macroscopic,yang1962concept}. It underlies phenomena ranging from laser and Bose-Einstein condensation to superfluidity and superconductivity. %~\cite{collins1960coherence,andrews1997interference,pereverzev1997quantum,josephson1962possible,frowis2018macroscopic}. 
Besides its fundamental importance, the ability to establish and maintain coherence among many particles is useful for quantum technologies including metrology, precise clocks, and quantum computing. Macroscopic coherence, however, is difficult to sustain in solids or under ambient conditions, where coupling to complex environments causes decoherence, while structural disorder and inhomogeneous broadening disrupt collective phase correlations~\cite{temnov2005superradiance,zurek2003decoherence}.

Superradiance provides a mechanism for generating macroscopic coherence of many photons by synchronizing many emitters into a collective dipole~\cite{dicke1954coherence,rehler1971superradiance,gross1982superradiance,scully2006directed, scully2009super,kocharovsky2017superradiance}. Superradiant emission has been observed in atomic and molecular ensembles~\cite{skribanowitz1973observation,gross1976observation,gross1979maser, norcia2016superradiance,kim2018coherent}, artificial emitters~\cite{van2013photon,mlynek2014observation,gottscholl2026semiconductor,wang2020controllable,raino2018superfluorescence}, and solids~\cite{scheibner2007superradiance,oxborrow2012room,raino2018superfluorescence, bradac2017room,angerer2018superradiant,xie2026observation, lei2023many,kersten2026self}. In most of these realizations, the stored excitation is released as a transient cooperative emission. Continuous pumping can sustain the collective coherence, producing superradiant lasing or masing~\cite{haake1993superradiant,meiser2009prospects,bohnet2012steady,jiang2021floquet,kraus2014room,baumann2010dicke,zhang2021observation}.
Continuous driving can take a collectively emitting system far from equilibrium, producing autonomous limit cycles or chaos~\cite{kessler2019emergent}. A robust limit cycle emerging under a time-independent drive can spontaneously break the continuous time-translation symmetry and form a continuous time crystal~\cite{iemini2018boundary,kongkhambut2022observation,liu2023photonic,greilich2024robust,huang2025observation,Wu2024Dissipative}. Most existing superradiant platforms, however, rely on dilute atomic gases, a relatively small number of emitters, cryogenic operation, or transient excitation. A large number of nitrogen-vacancy (NV) center spins in diamond can be optically polarized under ambient conditions, providing a compact platform for superradiant maser~\cite{jin2015proposal,wu2022superradiant,wu2026proposal,xiao2026squeezed}. Experiments have demonstrated room-temperature continuous-wave masers or amplifiers~\cite{breeze2018continuous,zollitsch2023maser,day2024room,sherman2022diamond,ohta2025near,ng2025portable}, cryogenic transient superradiance~\cite{angerer2018superradiant}, and cryogenic quasi-continuous superradiant masing~\cite{kersten2026self}. Nevertheless, sustaining superradiant coherence among a macroscopic number of solid-state spins under ambient conditions remains challenging.

Here we demonstrate a superradiant maser under ambient conditions, in which coherent oscillations of approximately $10^{14}$ nitrogen-vacancy spins and about $10^{9}$ intra-cavity microwave photons are synchronized and sustained. By varying the optical pump rate, cavity quality factor and spin-cavity detuning, we observed three distinct superradiant phases. Just above the threshold, the spins and photons synchronize to a fixed frequency and generate a continuous-wave (CW) superradiant maser. Well above the threshold, the collective spins develop a stable limit cycle, producing periodically amplitude-modulated (PAM) superradiant maser, which can be viewed as a coherent continuous time crystal. Deep above the threshold, the synchronization among collective spins breaks down and the maser exhibits superradiant bursts, which have random relative phases and intervals. These results show that macroscopic quantum coherence can be spontaneously generated and sustained in a solid under ambient conditions, and establish a platform for exploring non-equilibrium quantum many-body physics and bright light sources with collective quantum correlations~\cite{xiao2026squeezed}.

%\mynote{add remarks on limitations of existing works and why sustained (not pulsed) macroscopic (not few atoms) coherence in solids under ambient conditions is important.}

%\mynote{cite early superradiant laser and maser theory papers, including the recent one by  Klaus Molmer}

%\mynote{add review on THE recent work on rare-earth}

%\mynote{Review recent works on continuous time crystal works}. 

%\mynote{too many details about NV in introduction. shorten the paragraph}
%\mynote{revise this following the logic laid out in abstract}

%\section{Results}\label{sec2}

\section*{Setup and model for superradiant maser}\label{sec3}
\begin{figure}[htp]
    \centering
    \includegraphics[width=0.95 \textwidth]{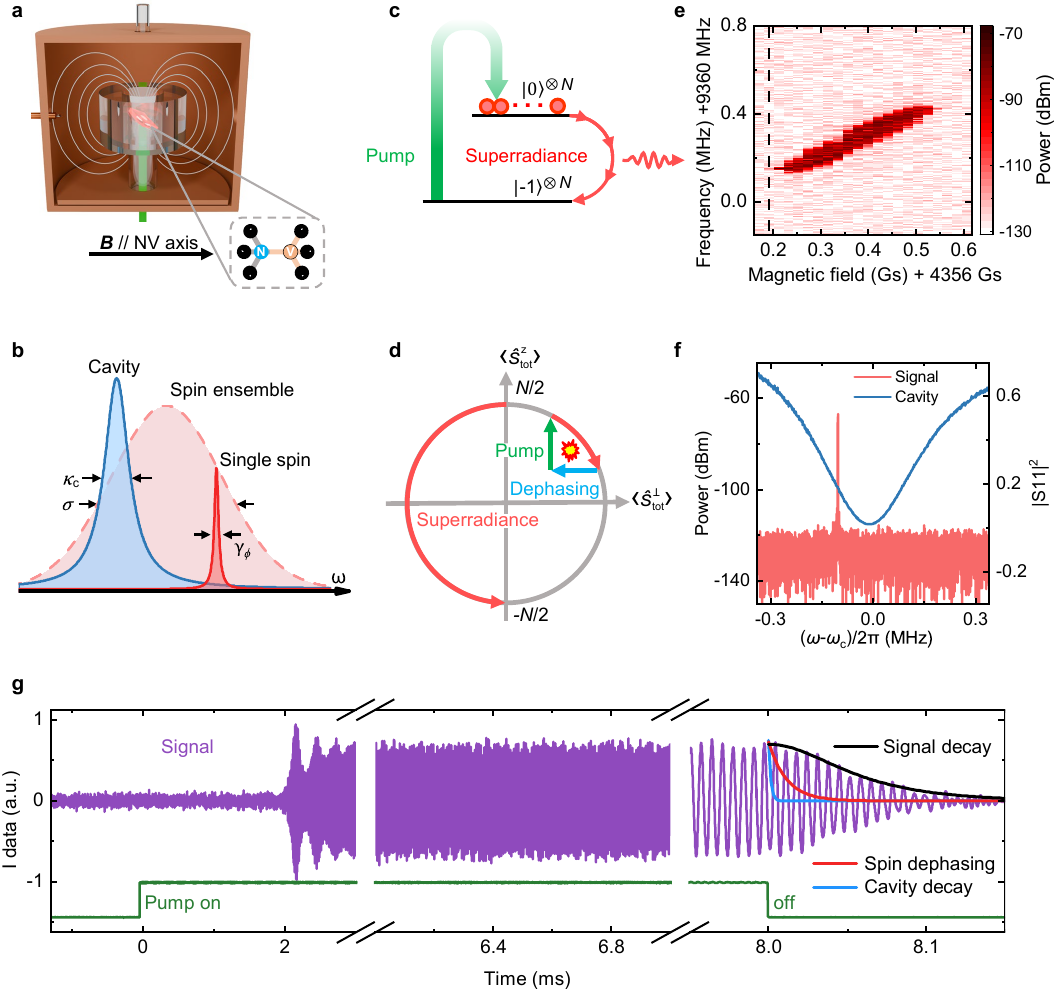}
    \caption{\textbf{Superradiant masing of a diamond under ambient conditions.}
    \textbf{a,} Setup. A diamond containing an ensemble of $\rm NV^{-}$  spins is placed inside a sapphire ring and coupled to a microwave cavity field (white lines, see also Extended Data Fig.~\ref{fig_s2}a). A magnetic field \emph{\textbf{B}} from an electromagnet, applied along the $\rm NV^{-}$  $\left\langle 111 \right\rangle$  axis, tunes the spin-transition frequency. The spins are optically pumped by a 532 nm laser, and the emitted microwave signal is detected by a loop antenna. Gas-flow cooling applied from above stabilizes the sample temperature (Method). 
    \textbf{b,} Schematic spectra of the cavity mode, the inhomogeneously broadened spin ensemble, and a single spin, with corresponding linewidth $\kappa_{\rm c}$, $\sigma$, and $\gamma_{\rm \phi}$.
    \textbf{c,} Level scheme of the superradiant process. Optical pumping polarizes the NV spins into the collective excited state $\left| 0 \right\rangle^{\otimes N}$, from which collective emission to the ground state $\left| - 1 \right\rangle^{\otimes N}$ generates superradiant maser signal.
    \textbf{d,} Superadiance by coherent rotation of the collective spin $\langle\hat{\mathbf S}_{\rm tot}\rangle$ from the north to the south pole (red semi-circle). %plane $(\langle\hat{S}_{\rm tot}^{\rm \perp}\rangle,\langle\hat{S}_{\rm tot}^{\rm z}\rangle)$  at the cavity frequency. 
    The balance among incoherent optical pumping (green upward arrow), spin dephasing (blue leftward arrow), and superradiance (red arc representing the spin rotation) gives rise to periodically modulated masing (the limit cycle formed by the three arrows), which approaches to steady masing at the zero-modulation limit (red spot). %  distinct superradiant trajectories (right). Without pumping and dephasing, the collective spin follows a circular trajectory (left) instead.
    \textbf{e,} Measured spectra of a superradiant maser as a function of the external magnetic field. The pump rate $W$ is 12 times the threshold $W_{\rm th}$, and the cavity quality factor $Q = 25,400$.
    \textbf{f,} Spectrum of a CW maser (red) corresponding to the dashed line in \textbf{(e)}, in comparison with the cavity reflection spectrum $\rm|S_{11}|^2$  (blue).
    \textbf{g,} Time-domain superradiant signal in response to optical-pump switching under the similar conditions as in \textbf{(f)}. The signal (purple line) stabilizes at 3 ms after the pump is switched on. When the pump is switched off, the coherent superradiance persists for more than 40~\textmu s, with envelope well fitted by a $\mathrm{Sech}$ curve (black line), much longer than the cavity decay time 0.66~\textmu s and the spin dephasing time $5.7$~\textmu s (exponential fits in blue and red lines, respectively). The carrier frequency is down shifted to 0.15 MHz for visualization.% More results showing superradiance after the pump is switched off are presented in Extended Data Fig.~\ref{fig_switchoff}.
    }
\label{Figure1}
\end{figure}

Figure~\ref{Figure1}a shows the setup. The NV spins within an isotopically purified diamond ($3\times 3\times 0.5 ~\rm~mm^3$) are coupled to the $\mathrm{{TE}_{01\rm \delta}}$ mode of a microwave cavity with a resonant frequency $\omega_{\rm c}/2\pi = 9.360$~GHz. The mode linewidth $\kappa_{\rm c}/2\pi$  is tunable from 0.23 MHz to 0.56 MHz, corresponding to a quality factor $Q$ from $17{,}000$ to 40,000. The diamond is mounted and aligned such that the [111] axis is along the external magnetic field applied in the horizontal direction. The NV centres have spin-1 in their ground state, with magnetic quantum number $m_{\rm s}=0$ and $\pm 1$. The magnetic field can be tuned to bring the transition frequency $\omega_{\rm s}$ between the spin states $\left|-1 \right\rangle$ and $\left|0 \right\rangle$ of the [111]-oriented NV centres into resonance with the cavity mode. The spins have a longitudinal relaxation rate  $\gamma/2\pi = 1/(2\pi T_1) = 37~\mathrm{Hz}$, an inhomogeneous broadening $\sigma/2\pi = 2/(2\pi T_2^*) = 0.42~\mathrm{MHz}$, and an average single-spin dephasing rate $\gamma_{\phi}/2\pi = 1/(2\pi T_2) = 0.056~\mathrm{MHz}$ (determined from the dephasing time under Hahn echo). See Extended Data Fig.~\ref{fig_s1} for the corresponding measurements and parameter extraction. The linewidth of the cavity mode, the inhomogeneous broadening of the spin ensemble, and the homogeneous broadening of a single spin due to dephasing are schematically shown in  Fig.~\ref{Figure1}b. 
A 532 nm laser is used to pump the NV spins  from the ground state $\left|-1 \right\rangle$  to the excited state $\left|0 \right\rangle$  to achieve population inversion (Fig.~\ref{Figure1}c). The spot size of the laser is approximately $3~\mathrm{mm}$ in diameter, corresponding to an estimated effective number of $N=1.40\times10^{14}$ optically pumped $\mathrm{NV}^{-}$ centres in resonance with the cavity mode (see Methods).
%\mynote{\textcolor{red}{In Methods, the estimation of this number should be described}}.
%\mynote{add information about the laser size and estimate the number of NV spins within the laser spot. with an effective number of NV$^-$  centres $N = 1.44 \times 10^{14}$.}
 The optical pump rate $W$ is tuned by rotating a half-wave plate to change the pump-laser polarization relative to the NV axis (with the heating effect of the laser kept nearly constant).
%\mynote{add description of how W is tuned}. 
The emitted microwave signal is coupled out of the cavity through a loop antenna and recorded with a spectrum analyzer (Fig.~\ref{Figure1}a).
Superradiance occurs when $W$ exceeds a threshold  {(which depends on the cavity $Q$ factor and the spin-cavity detuning $\Delta\equiv \omega_{\rm s}-\omega_{\rm c}$)}.
%\mynote{\textcolor{red}{I added this. Make sure the notations are consistent with your definitions}}.

This spin-cavity system is described by the Hamiltonian ($\hbar$ set as unity)%\mynote{I added $G$ to $g$ in the last term. Check it. Checked and updated accordingly.}
\begin{equation}
    H = \omega_{\rm c}\hat{a}^{\dagger}\hat{a} + \sum_{j=1}^{N}\omega_j\hat{s}_j^z + i g(\hat{a}^{\dagger}\hat{S}_{\rm tot}^- - \hat{a}\hat{S}_{\rm tot}^{+}),
    \label{Hamiltonian}
\end{equation}
where $\hat{s}_j^z$ is the $j$-th spin-1/2 for the transition between the states $\left|-1 \right\rangle$  and $\left|0 \right\rangle$ and the collective spin $\hat{S}^{z/+/-}_{\rm tot}\equiv\sum_{j=1}^N\hat{s}^{z/+/-}_j$, with the lower operator $\hat{s}^{-}_j\equiv |-1\rangle\langle 0|$ and the raiser operator $\hat{s}^{+}_j\equiv |0\rangle\langle -1|$. {The  frequency $\omega_j$ of the $j$-th spin is Gaussian-distributed around the central frequency $\omega_{\rm s }$ with a standard deviation $\sigma/\sqrt{2}$ (see Supplementary Information \ref{SI1}).} 
The spin-cavity coupling is approximated as uniform with a coupling constant $g$. The measured collective spin-cavity coupling strength is $G/2\pi \equiv \sqrt{N} g/{2\pi}= 0.32\text{~MHz}$ (see Extended Data Fig.~\ref{fig_s2}c).
%\mynote{add cross-reference to extended figure or methods}).
The superradiance is the radiation of photons resulting from the coherent rotation of the collective spin, as illustrated by a collective Bloch vector in Fig.~\ref{Figure1}d. The coherent collective spin–cavity interaction converts the longitudinal population inversion into macroscopic transverse spin coherence $\langle\hat{S}_{\rm tot}^{\perp}\rangle$  (curved arrow) and photon coherence. On the other hand, the spin dephasing  (blue arrow) suppresses the transverse coherence, bringing the collective Bloch vector toward the $z$-axis. The incoherent optical pumping   (green arrow) replenishes the spin excitation and increases the longitudinal component $\langle\hat{S}_{\rm tot}^{z}\rangle$. The balance among the optical pumping, the superradiance, and the spin dephasing can in general sustain a cyclic dynamics of the macroscopic spin-photon system, which in the limiting case becomes a fixed point (the flash spot in Fig.~\ref{Figure1}d).

Figure~\ref{Figure1}e shows the superradiant maser spectrum as a function of the external magnetic field for a fixed cavity quality factor $Q=25{,}400$ and the pump rate $W/W_{\rm th}=12$. As the spin-transition frequency is tuned away from the cavity frequency, the spectral linewidth progressively narrows. Owing to the frequency dragging effect, the masing frequency 
% \deleted{$\omega_{\rm m}=(\kappa_{\rm s}\omega_{\rm c}+\kappa_{\rm c}\omega_{\rm s})/ (\kappa_{\rm c}+\kappa_{\rm s})$ (where $\kappa_{\rm s}= $...\mynote{add definition})} 
% I removed it because the dragging effect has no analytical expression in the inhomogeneous model, while using the homogeneous-model result would make the argument logically inconsistent.
is drawn towards the cavity resonance: varying the magnetic field by $0.35~\mathrm{Gs}$ shifts the spin-transition frequency by $0.98~\mathrm{MHz}$, but the maser frequency by only $0.27~\mathrm{MHz}$. Fig.~\ref{Figure1}f shows the continuous-wave (CW) superradiant spectrum for spin-cavity detuning just satisfies the above-threshold condition (marked by dashed line in Fig.~\ref{Figure1}e). The maser frequency is $0.10~\mathrm{MHz}$ shifted away from the cavity mode, which is much smaller than the spin-cavity detuning $0.50~\mathrm{MHz}$ due to the frequency dragging in superradiance. The CW emission has an output power of $P_{\rm out}\simeq-67~\mathrm{dBm}$ and a full width at half maximum (FWHM) of $128~\mathrm{Hz}$. The maximum CW superradiant output power measured in our experiment is $P_{\rm out}^{\max}\simeq-52.3~\mathrm{dBm}$, corresponding to a number of intra-cavity photons about $1.45 \times 10^9$, under an optical pump power of $1.7~\mathrm{W}$ (Extended Data Fig.~\ref{fig_maxPower}).

To check whether the masing is due to superradiance, we measure the signal after the pump laser is suddenly switched off.  Figure~\ref{Figure1}g shows the time-domain signal of a typical CW superradiant maser. %For visualization, the signal is down-converted to a frequency of 0.15 MHz. 
When the pump laser is switched on (green), the population inversion starts to build up. The maser signal appears at $\sim$ 2~ms, followed by a steady continuous-wave signal. When the pump is switched off, the coherent emission remains for more than $40$~\textmu s, with a profile well-fitted by the characteristic Sech curve of superradiance (black line). Remarkably, the decay time of the emission is far exceeding the cavity-photon lifetime ($0.66$~\textmu s,  blue line) and the spin dephasing time ($T_2=5.7$~\textmu s, red line), but much smaller than the  relaxation time of the spins ($T_1\approx 4.3~\rm ms$). This large separation of timescales indicates that the coherence is stored in the spin ensemble rather than in the cavity field. Superradiance suppresses not only the effects of inhomogeneous broadening but also spin dephasing through coherence synchronization (frequency dragging). A more systematic study of residual emission after pump switch off is shown in Fig.~\ref{fig_switchoff} as evidence of the superradiant nature of the diamond maser.

\section*{Synchronization of macroscopic quantum coherence}\label{sec4}

\begin{figure}[htp]
    \centering
    \includegraphics[width=0.95 \textwidth]{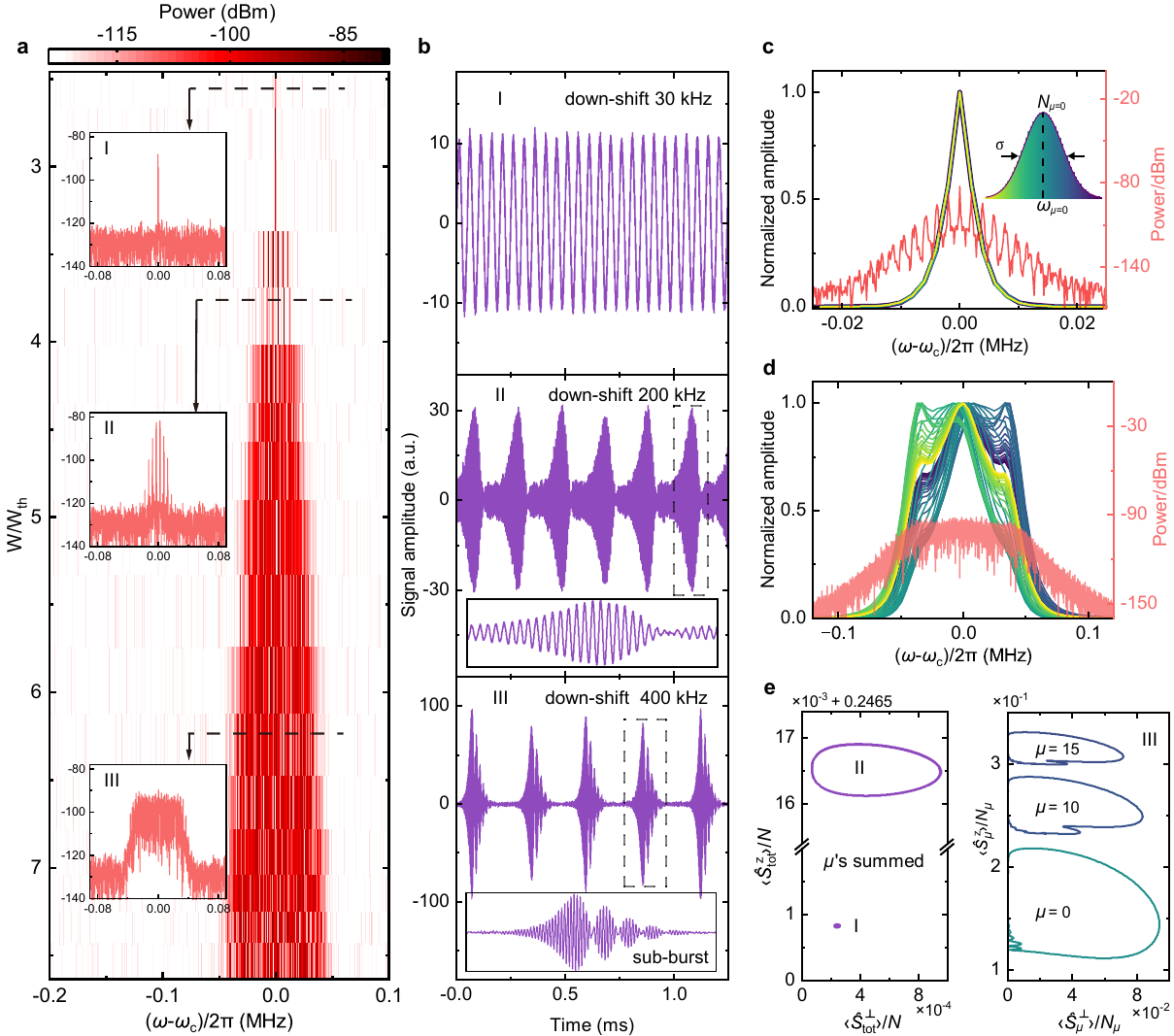}
    \caption{\textbf{Synchronization of macroscopic coherence in a superradiant diamond.}
    \textbf{a,} Spectrum of the superradiant maser as a function of pump rate. The spin-cavity detuning is zero and the cavity quality factor $Q\simeq 24,580$.  The insets show representative spectra extracted from dashed lines.
    \textbf{b,} Time-domain signals corresponding to spectra I-III in \textbf{(a)}, showing, from top to bottom, CW, periodic amplitude modulation, and bursts. For clarity, the carrier frequencies are downshifted to $30~\mathrm{kHz}$, $200~\mathrm{kHz}$ and $400~\mathrm{kHz}$, correspondingly; the dashed-boxed regions are enlarged insets.
    \textbf{c,} Numerically simulated spectra of the spin and photon coherence in the regime of periodic amplitude modulation. Inset shows the frequency distribution of the spin ensemble with inhomogeneous broadening $\sigma/2\pi= 0.42$~MHz, which is divided into 321 sub-ensembles of similar transition frequencies. The normalized spectra of the 321 spin sub-ensembles, obtained by Fourier transform of the coherent spin oscillations in a single period, are almost identical (colored curves; left axis). The collective emission spectrum evaluated over the full pulse sequence forms a frequency comb (red; right axis).
    \textbf{d}, Similarly to  \textbf{(c)}, but obtained by simulation in the regime of superradiant bursts. The spectra of the 55 representative sub-ensembles in a single burst are distributed in a range (with line colors corresponding to the colors of spin-sub-ensembles in inset of \textbf{(c)}). %, with the central sub-ensembles exhibiting the largest frequency dispersion and the peripheral sub-ensembles remaining approximately synchronized. 
    The collective radiation spectrum exhibits a plateau (red; right axis).
    \textbf{e}, The trajectories of spin sub-ensembles in collective Bloch sphere, computed with parameters corresponding to the cases I, II and III in \textbf{(a)}. In the CW regime ({I}), all spin sub-ensembles converge to fixed points, collectively represented by a fixed point. In the PAM regime ({II}), all spin sub-ensembles follow closed trajectories with the same period and phase, collectively forming a limit cycle. % (as summed over sub-ensemble index $\mu$). 
    In the burst regime ({III}), the spin sub-ensembles follow closed trajectories with different shapes, periods, and phases. These trajectories can  include a segment along the $z$-axis where the transverse spin polarization (i.e., the macroscopic spin coherence) vanishes, as illustrated for the sub-ensembles with index $\mu=0$, $10$ and $15$.
    In the simulations for \textbf{(c-e)}, the parameters are chosen similar to experiments in \textbf{(a)}, with the longitudinal spin-relaxation rate $\gamma/2\pi=37~\mathrm{Hz}$, spin dephasing rate $\gamma_\phi/2\pi=0.056~\mathrm{MHz}$, collective spin-cavity coupling strength $G/2\pi=0.32~\mathrm{MHz}$ and a mean thermal photon number $\bar{n}_{\mathrm{th}}=670$ within the cavity {linewidth}. The pump rates in the simulations are $W/W_{\rm th}=2.21$, $2.54$, and $7.36$ for I, II, and III in turn.
    % \added{The pump-rate ranges of phases I and II differ slightly from the experimental results, possibly owing to deviations from the assumed Gaussian spin-frequency distribution and uncertainties in pump-rate calibration.}
    }
    \label{Figure2}
\end{figure}

Figure~\ref{Figure2}a presents a spectrum of the maser as a function of pump rate $W$ for a fixed cavity quality factor ($Q=24{,}580$, measured with the diamond loaded) and spin-cavity resonance ($\Delta=0$). As the pump rate increases, the spectrum evolves from a single sharp peak to a frequency comb and eventually to a plateau-shaped spectrum. The insets labeled I, II, and III show three representative cases, with the corresponding time-domain signals shown in Fig.~\ref{Figure2}b exhibiting in turn CW maser, PAM, and well-separated bursts. 

To understand the physical origins of the three different regimes, we perform numerical simulations in the framework of mean-field theory based on the Hamiltonian in equation~\eqref{Hamiltonian} including the effects of cavity leakage, spin relaxation and dephasing, and incoherent pump (see Methods for details). To account for the large inhomogeneous broadening of the spin ensemble, we group the spins into 321 sub-ensembles, each of which has almost the same frequency (inset of Fig.~\ref{Figure2}c). The frequency and the number of spins in the $\mu$-th sub-ensemble are denoted as $\omega_{\mu}$ and $N_{\mu}$, respectively. Thus the maser is regarded as the superradiance of 321 collective spins into a single cavity mode. Each individual spin sub-ensemble, if coupled independently to a cavity mode, would have their distinct threshold and maser frequency due to their distinct detuning from the cavity mode. However, when coupled to a common cavity mode, the spin sub-ensembles present coherent oscillation synchronized by the frequency dragging effect in sustained superradiance.

In the CW maser case (I in Fig.~\ref{Figure2}b), the emitted microwave signal oscillates with a nearly constant amplitude envelope, and its spectrum presents a single sharp peak with an FWHM of 164 Hz (I in Fig.~\ref{Figure2}a), indicating a coherence time of 1.94~ms. In the simulation with parameters similar to those in experiments, we find that 
all spin sub-ensembles are driven to fixed points in the reference frame rotating with the maser frequency  (Fig.~\ref{Figure2}e, I), i.e., all sub-ensembles are synchronized to the maser. Note that the summed fixed point responds over all sub-ensembles has a non-zero transverse component, that is, $\left\langle \hat{S}_{\rm tot}^{\perp}\right\rangle\sim O\left(N\right)$, which means sustained macroscopic spin coherence.
   
The maser enters the PAM regime under a moderate pump (e.g., II in Fig.~\ref{Figure2}b, {with a modulation period of 0.2~ms}). Correspondingly, the spectrum presents a frequency comb (II in Fig.~\ref{Figure2}a, with a spacing of $5~\mathrm{kHz}$ {between sharp lines with FWHM $\sim 145$~Hz}). The PAM can be understood as an effect of periodic trajectories of the large spin associated with the spin sub-ensembles. Under pumping stronger than that in the CW maser regime, the spin sub-ensembles can be pumped to larger spin states, which would have faster (and stronger) superradiance. Then there could be a segment of time in which the spin dephasing is not fast enough to balance the superradiance (red arrow in Fig.~\ref{Figure1}d), leading to a limit cycle of the collective Bloch vector and hence the PAM. This phenomenon is indeed reproduced in the numerical simulation. Under a condition  similar to the experiment (II Figs.~\ref{Figure2}a,b), the spin sub-ensembles all form cyclic trajectories on the collective Bloch sphere. The cycles of all sub-ensembles are synchronized with the same period and phase, which is evidenced by the fact that their normalized spectra  all overlap (Fig.~\ref{Figure2}c, {see Supplementary Information Fig.~\ref{fig_s4}b for time-domain data)}. Correspondingly, the maser spectrum (Fig.~\ref{Figure2}c) presents a frequency comb similar to the experimental data. 
Note that in the whole cycle the macroscopic spin coherence is sustained, i.e., $\left\langle \hat{S}_{\rm tot}^{\perp}(t)\right\rangle\sim O\left(N\right)$ (see the summed trajectory of all spin sub-ensembles, II in Fig.~\ref{Figure2}e).  The spin coherence can seed the superradiance in each cycle, making the macroscopic coherence sustained and synchronized. As a result, the photon coherence oscillation persists between peaks (as observed in experiments, see case II of Fig.~\ref{Figure2}b). Such self-sustained oscillation of macroscopic spin and photon coherence spontaneously breaks the continuous time-translation symmetry of the system with non-trivial off-diagonal large-scale coherence, realizing a quantum-coherent continuous time crystal in a solid under ambient conditions. 

Under strong pumping, the emission presents a plateau without resolved spectral lines (III in
Fig.~\ref{Figure2}a) and, in the time-domain, well-separated multi-peak bursts without coherent oscillations in between (III in Fig.~\ref{Figure2}b). The plateau-like spectrum indicates that there is no phase coherence between the bursts. The numerical simulation shows that the dynamics of different spin sub-ensembles in a single burst have different spectra -- only the peripheral sub-ensembles (those far detuned from the cavity resonance) are approximately synchronized (Fig.~\ref{Figure2}d).
In addition, the collective Bloch vector trajectories of different sub-ensembles in general have different periods, shapes, and phases (except for those peripheral ones), as illustrated by the representative cases for the $0$-th, $10$-th, and $15$-th sub-ensembles (III in Fig.~\ref{Figure2}e, {see Supplementary Information Fig.~\ref{fig_s4}e for time-domain data}).  Remarkably, in the collective spin trajectories there are segments
along the $\langle\hat{S}_{\mu}^{z}\rangle$  axis, where the transverse polarization and the superradiant emission almost vanish, which means absence of the macroscopic coherence. {The existence of such incoherent segments can be understood from the high polarization of the central sub-ensembles (the ones near resonant with the cavity mode) under a strong pump, which enables a small number of nearby sub-ensembles to form a collective spin that super-radiates independently and fast enough to complete before the pump can replenish the spin polarization for sustained coherence.
During these segments, incoherent pumping repolarizes the spin ensemble until the next superradiant burst is triggered spontaneously.} Therefore, there is a lack of phase correlation or synchronization between the neighboring superradiant bursts (hence the plateau-like spectra, see Fig.~\ref{Figure2}d and inset III in Fig.~\ref{Figure2}a). In addition, the waiting times (the intervals between bursts) and the burst amplitudes fluctuate, with higher amplitudes correlated with longer waiting times as expected (see Extended Data Fig.~\ref{fig_interval_amp_corr}).

\section*{Residual emission after pump switched off}

\begin{figure}[H]
\centering
\includegraphics[width=0.95\textwidth]{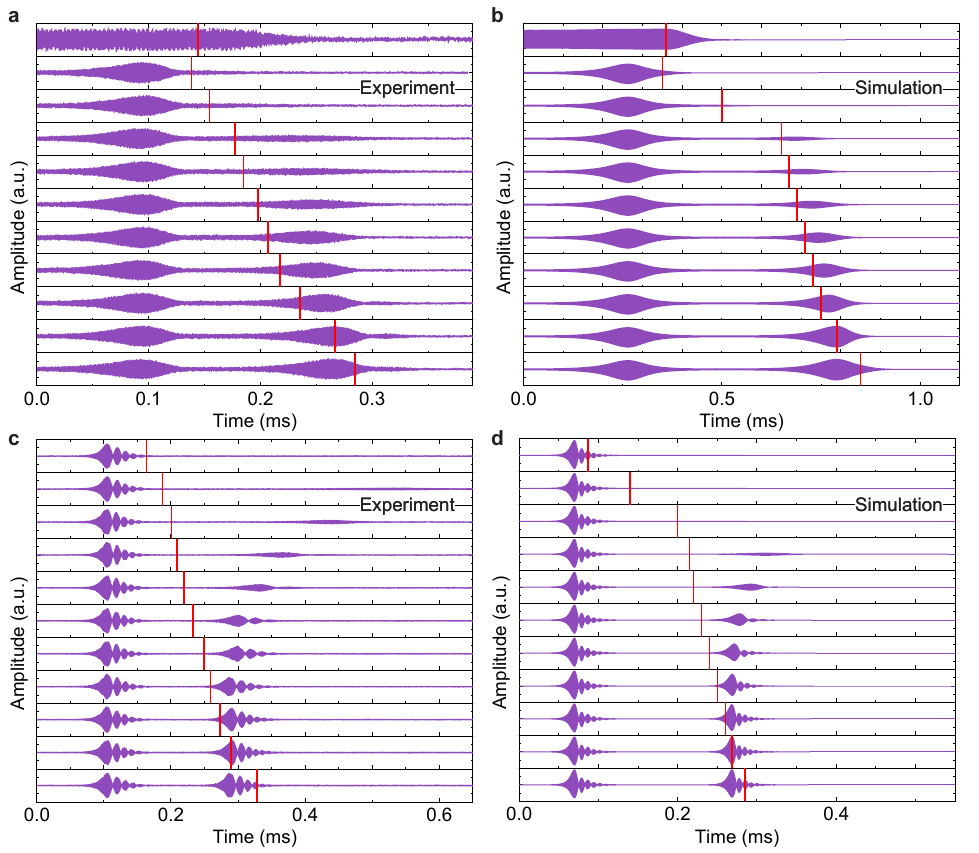}
\caption{\label{fig_switchoff}
\textbf{Microwave emission after pump is switched off.} The red bars mark the time when the laser is switched off. 
 \textbf{(a/b)} are experimental/numerical results in the CW (first row) and PAM (other rows) regimes. \textbf{(c/d)} are the experimental/numerical results in the burst regime.  
The pump rates for the CW, PAM, and burst cases are in turn $W/W_{\rm th}=2.21$, $2.54$, and $13.4$.  The spin-cavity detuning $\Delta=0$. All other parameters are the same as in Fig.~\ref{Figure2}.% A reference oscillation with $\omega_{\rm r}=2$~MHz is added to the simulated signal for comparison.
}
\end{figure}

We measure the residual emission after the pump is suddenly turned off (Fig.~\ref{fig_switchoff}). In all of the three maser regimes, the emission persists for a duration ($40$-$50$~\textmu s) much longer than the cavity photon leakage time $1/\kappa_{\rm c}$ and the spin dephasing time $T_2$, but much shorter than the spin relaxation time $T_1$. In the PAM and burst cases, if the pump is switched off at the peak of an emission cycle, the emission persists with its amplitude, shape, and phase nearly unchanged until the cycle is completed; if the pump is switched off at the end of an emission cycle (when the spin polarization is nearly exhausted to sustain superradiance),  the residual emission has a small amplitude in the PAM regime and nearly vanishes in the burst regime; if the pump is {switched off} during the interval (so that the spin polarization has been built up), the residual emission ramps up and then decays, with shorter waiting time and larger peak amplitude for longer delay between the switch-off and the completion of the previous cycle. These measured features are well reproduced by numerical simulations under similar conditions. The features of the residual emission after the pump switch-off and their dependence on the pump power and switch-off timing confirm the superradiant nature of the emission and are consistent with the fixed or cyclic dynamics of the collective spins of the sub-ensembles.

\section*{Phase diagram}

\begin{figure}[!htbp]
    \centering
    \includegraphics[width=0.95 \textwidth]{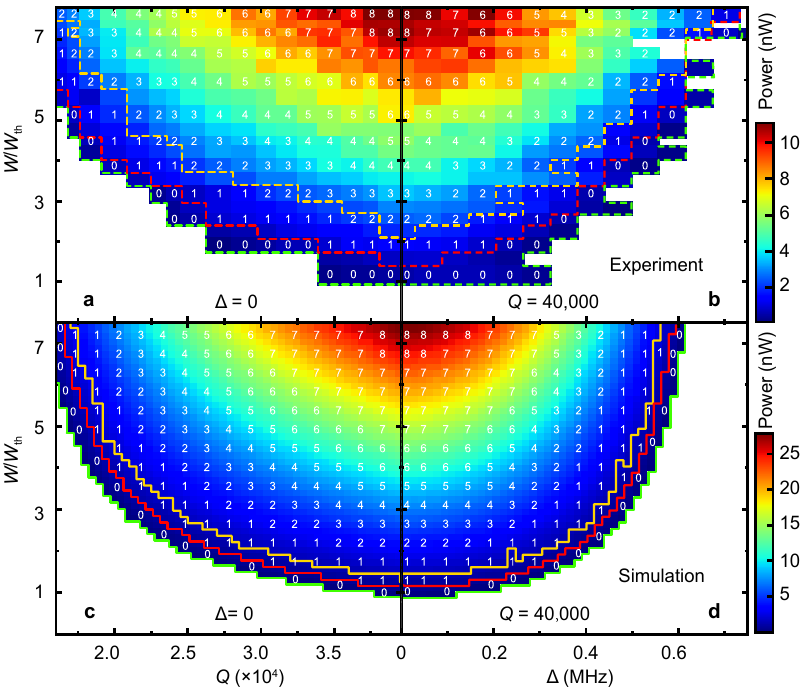}
    \caption{\label{fig:phase_diagram}
    \textbf{Phase diagram of the superradiant diamond maser.}
    \textbf{a,} Measured intra-cavity superradiant power (color map) and number of peaks per burst/period (white integers) versus pump rate {(normalized by the threshold pump rate $W_{\rm th}$ at  $Q=40,000$ and $\Delta=0$)} and cavity quality factor under the spin-cavity resonance condition (detuning $\Delta=0$). Dashed lines divide the superradiant phase diagram into phases: CW maser (between the green and red dashed lines); PAM superradiance with a frequency-comb spectrum (between the red and yellow dashed lines); and superradiant burst with a plateau spectrum (above the yellow dashed line).
    \textbf{b,} Same as \textbf{(a)}, but as a function of the pump rate and the spin–cavity detuning $\Delta$ for a fixed cavity quality factor ($Q=40{,}000$).
    \textbf{c} and \textbf{d}, Same as \textbf{(a/b)}, but obtained by numerically simulation.
   All parameters used in the numerical simulations are the same as in Fig.~\ref{Figure2} if not otherwise specified.
}  
\end{figure}

We identify three phases of the superradiant maser, the CW phase, the PAM phase which can be regarded as a coherent continuous time crystal, and the incoherent burst phase for pump, correspondingly, above, well above, and deep above the threshold. When the pump rate is fixed but the spin-cavity detuning is varied, similar phases are also observed under corresponding above threshold conditions (Extended Data Fig.~\ref{fig_s9}).
We map the boundaries of these phases by tuning the pump rate, the cavity quality factor and the spin-cavity detuning (Fig.~\ref{fig:phase_diagram}). The experimental data and the numerical simulations with similar parameters are in good agreement.

%\section{Phase diagram of the superradiant maser}\label{sec5} 
% The superradiant phases are governed by several parameters, including the pump rate, cavity quality factor, spin-cavity detuning and number of participating NV centres. This raises the question of how the boundaries between different superradiant phases distributed. We therefore obtain experimental and theoretical phase diagrams of the superradiant maser signals. 

Under the spin-cavity resonance condition ($\Delta=0$), we map the intra-cavity superradiant power, $P=P_{\rm out}\kappa_{\rm c}/\kappa_{\rm in}$, as a function of the pump rate $W/W_{\rm th}$ and cavity quality factor $Q$ (Figs.~\ref{fig:phase_diagram}a,c). The intra-cavity power reaches a maximum of $12~\mathrm{nW}$ with a pump power of $256~\mathrm{mW}$, and the minimum pump rate, $W/W_{\rm th} = 1$, corresponds to a pump power of $33~\mathrm{mW}$. The threshold pump power is approximately an order of magnitude lower than previously reported values~\cite{breeze2018continuous,zollitsch2023maser}, owing to the larger cooperativity of our system.  The three phases (boundaries shown as dashed lines) {are identified by whether the spectra exhibit a frequency-comb structure}.
We also mark the number of envelope peaks in each period/burst on the phase diagrams (white integers), assigning $0$ to the CW phase. It is noted that the peak count is more than one in the superradiant burst phase and increases with increasing pump. This increase of peak count per burst is consistent with the picture that under stronger pumping the spin sub-ensembles can be polarized to a higher degree and therefore form more groups of sub-ensembles which super-radiate independently. %Indeed, under a smaller pump  and higher pump power, the bursts with higher sub-burst counts can still exhibit a well-resolved frequency comb and maintain strong collective coherence.
Note that the multi-peak structure can be observed even in the PAM phase (see, e.g., Extended Data Fig.~\ref{fig_partial_synchronization}), which indicates that the spin sub-ensembles are not fully synchronized but are self-organized into several individually synchronized groups, whose interference leads to the multi-peak structure (see {Supplementary Information}~\ref{SI4} and Fig.~\ref{fig_s4}e for confirmation by numerical simulation).

For a fixed $Q = 40{,}000$ , we map the intra-cavity superradiant power as a function of the pump rate and the spin-cavity detuning (Figs.~\ref{fig:phase_diagram}b,d). The phase diagrams are similar to those in Figs.~\ref{fig:phase_diagram}a,c considering how much the parameters are above the maser threshold. %Different from the resonance case, the spin-cavity detuning breaks the symmetry between the spin sub-ensembles with respect to the, resulting in incomplete destructive interference to the cavity field, as manifested by the non-zero signal minima between successive sub-bursts (Extended Data Fig.\ref{fig_s9}).

We also show the period (or average interval in the burst phase) in Extended Data Fig.~\ref{fig_s7}. In both the PAM phase and the burst phase, the period or burst interval decreases with the pump power, which is understood since stronger pump makes the time needed to build up spin polarization for superradiance shorter and the superradiance faster (due to higher polarization). Interestingly, there is a non-monotonic increase of the period/interval at the boundary between the two phases. This is consistent with the physical picture that in the burst phase, the whole spin ensemble will be self-organized into several groups each superradiating independently. Since the number of spins in each group is smaller than the whole ensemble, it would need more time to build spin polarization to reach the superradiance threshold.

\section*{Conclusion and Outlook}\label{sec6}

We have demonstrated sustained macroscopic coherence among approximately $10^{14}$ NV spins and $10^{9}$ intra-cavity microwave photons in a superradiant diamond maser under ambient conditions. The interplay among optical pumping, spin dephasing and spin-cavity coupling gives rise to three distinct phases: continuous-wave superradiance, periodically amplitude-modulated superradiance forming a coherent time crystal, and superradiant bursts without mutual phase correlation. These phases can be controlled via the pump rate, cavity quality factor and spin-cavity detuning. Importantly, the inhomogeneous spin-frequency distribution not only causes dephasing but also shapes the collective dynamics. Our results establish that macroscopic quantum coherence can be spontaneously generated, controlled and sustained in a solid under ambient conditions.

This superradiant solid provides a platform for exploring nonequilibrium quantum many-body dynamics and bright quantum microwave radiation. Engineering direct and cavity-mediated spin interactions could generate many-body correlations within the ensemble~\cite{lewis2018robust,luo2025hamiltonian}, which could subsequently be imprinted on the emitted microwave field through collective superradiant emission~\cite{haake1996quantum,andersen2012squeezing,xiao2026squeezed}. More broadly, engineering the spin-frequency distribution and collective interactions could provide access to higher-order many-body correlations, with potential applications in quantum-enhanced sensing and metrology, quantum communications, and quantum nonlinear spectroscopy.

After completion of this manuscript, we notice that similar results are reported in a preprint by Zollitsch et al~\cite{Zollitsch2026DiamondMaser}.

\backmatter

%\bmhead{Supplementary Information}

%If your article has accompanying supplementary file/s please state so here. 

%Authors reporting data from electrophoretic gels and blots should supply the full unprocessed scans for key as part of their Supplementary information. This may be requested by the editorial team/s if it is missing.

%Please refer to Journal-level guidance for any specific requirements.

\bmhead{Acknowledgments}
 We thank {Xiao-Bing Liu and Xiao-Ran Zhang for providing NV-diamond samples for preliminary tests,} 
 Chong Chen for helpful discussions and comments, and Shuai-Wei Guo, Kang-Yuan Liu, Hong-Yu Pei, Shuo Wang, Jun-Chen Ye, Guo-Li Zhu, Yao Gao and Run-Ze Liu for discussions on NV charge-state conversion, temperature-dependent effects, Raman spectroscopy, and data cross-checking.

\bmhead{Author contributions}
R.B.L. conceived and supervised the project, W. J. W. and W.T.W. constructed the setup and carried out the measurements,  D.W.X. and R.B.L. developed the theoretical description, D.W.X. carried out the simulation, W. J. W., W.T.W., D.W.X., and R.B.L. analyzed and interpreted the data. X.Y.C., X.F.W., M.Z.A., and Q.L. contributed to the experimental design, characterization of the NV centres and the diamond sample preparation. W.J.W., D.W.X., W.T.W and R.B.L wrote the manuscript and all authors commented on it. 

\bmhead{Funding}
 This work was supported by the New Cornerstone Science Foundation, the National Natural Science Foundation of China/Hong Kong Research Council Collaborative Research Scheme Project CRS-CUHK401/22, and the Hong Kong Research Grants Council Senior Research Fellow Scheme Project SRFS2223-4S01.

\section*{Declarations}

%Some journals require declarations to be submitted in a standardised format. Please check the Instructions for Authors of the journal to which you are submitting to see if you need to complete this section. If yes, your manuscript must contain the following sections under the heading `Declarations':

%\begin{itemize}
The authors declare no Conflict of interest/Competing interests.
%\item Ethics approval and consent to participate
%\item Consent for publication
%\item Data availability 
%\end{itemize}

%\noindent
%If any of the sections are not relevant to your manuscript, please include the heading and write `Not applicable' for that section. 

%\begin{appendices}
\setcounter{figure}{0}
\renewcommand{\thefigure}{A\arabic{figure}}
\renewcommand{\theHfigure}{suppfigure.\arabic{figure}}

\section*{METHODS}\label{secA1}

\subsection*{Cavity} The cavity is composed of a cylindrical copper cavity, a sapphire ring resonator, and a hollow cylindrical quartz support. The copper cavity has an inner diameter of 14~mm and a height of 20~mm. A loop antenna is inserted through a side hole in the cavity to detect the signal. Another hole on the top connects a pipe to an exhaust system, which allows gas to flow through and dissipate the heat generated by the diamond. Without gas flow, high pump power would lead to a significant increase of temperature, which would substantially shorten the spin relaxation and dephasing times. As a result, the effective cooperativity would fall below the masing threshold, and the emission signal would disappear. The resonator has a height of 6~mm, an inner diameter of 2.5~mm, an outer diameter of 5.2~mm, and a relative permittivity of 9.4. We use the TE$_{01\rm{\delta}}$ mode, which has resonant frequency  around 9.360~GHz, in agreement with the COMSOL simulation (Extended Data Fig.~\ref{fig_s2}a,b).

\subsection*{Diamond sample and spin properties} The diamond is a single crystal of size $3\times 3\times0.5$~mm$^3$ from Element Six with $^{12}\mathrm{C}$ purified to a concentration 99.99\%, grown on the \{001\} surface by chemical vapor deposition. According to the manufacturer’s specifications, the initial nitrogen concentration [N] is 13~ppm and the negatively charged NV concentration [NV$^-$] is approximately 4.5~ppm after treatment. Using the XY8-$N$ dynamical decoupling sequence~\cite{zhang2026unraveling}, we measure the [NV$^-$] concentration to be around $2.7\pm0.1$~ppm (Extended Data Fig.~\ref{fig_s1}f,~g). {The number of [NV$^-$] centres effectively pumped and coupled to the cavity mode is estimated to be $N\approx N_{\rm NV}(\pi/4)(1/4)(1/3)\approx (1.40\pm0.05)\times10^{14}$, accounting for the fraction of volume illuminated by the pump laser with a spot diameter 3~mm ($\pi/4$), the fraction of NV centres along the magnetic field direction ($1/4$), and the $^{14}$N hyperfine state ($1/3$).} %When the sample was first inserted into the cavity, the loaded quality factor was so low that no discernible resonance dip was observed in the $\rm{|S_{11}|^2}$ reflection spectrum. 
To remove possible graphitic carbon from the surface, the sample has been treated in a mixture of sulfuric acid, nitric acid, and perchloric acid at $210^\circ\mathrm{C}$ for 3 hours, and subsequently cleaned by plasma immersion ion implantation with a nitrogen–oxygen mixture, and finally surface polished. The loaded $Q$ factor is up to 40,000. In situ measurements yields a longitudinal spin-relaxation time $T_1= 4.3~ \rm{ms}$, a Hahn-echo coherence time $T_2=5.7~ \rm{\mu s}$, an inhomogeneous dephasing time $T_2^*=754~\rm ns$, a magnetic field inhomogeneity of $< 0.064~ $Gs across the diamond, and a collective spin-cavity coupling strength $G/2\pi= 0.32\,\rm{MHz}$ (Extended Data Fig.~\ref{fig_s1} and ~\ref{fig_s2}c). 

\subsection*{Superradiant maser generation and characterization} 
 A tunable electromagnet provides the magnetic field along the horizontal direction. To align the diamond [111] axis with the horizontal external magnetic field, we use a custom quartz support with an inclined surface at $35.3^\circ$. A continuous 532 nm laser (CNI MGL-F-532) applied from the bottom of the diamond pumps the spins. The effective optical pump rate is controlled by rotating the laser polarization with fixed laser power, so that the temperature increase due to laser heating is kept constant for different pump rates. Rotating the polarization from perpendicular to parallel to the {NV$^-$} axis allows the effective pump rate to be tuned from its maximum value to nearly zero.  
 %The pump rate is estimated from $W = {\sigma_{\rm a} P}/{(\hbar \omega_{\mathrm{p}} A)}$, where $\sigma_{\rm a} = 3.1 \times 10^{-17}~\mathrm{cm}^2$ is the {reported} one-photon absorption cross-section of NV centres at 532 nm~\cite{wee2007two}, $P$ is the effective laser power (with the Fresnel reflection loss at the diamond surface taken into account), $\omega_{\mathrm{p}}$ is the laser angular frequency, and $A$ is the area of the laser spot on the diamond. Because of the large uncertainty in the cross section, this estimated $W$ serves only as an approximation, but the ratio of $W$ to the threshold $W_{\rm th}$ is not affected by the uncertainty of the cross section.

The microwave signal is coupled out through a loop antenna and routed to external measurement instruments. The time-domain signal is measured using an oscilloscope (SDS5014X) in both the I and Q channels, and the spectra are measured using a spectrum analyzer (thinkRF R5550). The loaded $Q$ factor is tuned by inserting a pencil lead into the cavity to increase the conductive loss. The $\left|S_{11}\right|^2$ measurement using a vector network analyzer (VNA: KEYSIGHT P9373A) characterizes the $Q$ factor and anti-crossing due to spin-cavity coupling. We record signals in both the time and frequency domains while sweeping the magnetic field for different pump rates and $Q$ factors to obtain the phase diagrams. The magnetic field sweep range is set to be wide enough to cover all detunings where the superradiant signal appears.

\subsection*{Superradiant maser model}
%The observed superradiant emission and synchronization dynamics can be explained by an inhomogeneously broadened single-mode superradiant maser model, which qualitatively accounts for the observed phenomena and reproduces the quantitative trends. 
The model consists of $N$ spin-$1/2$ emitters coupled to a single cavity mode, and its dynamics is governed by the master equation~\cite{carmichael2013statistical}
\begin{align}
\frac{d\hat\rho}{dt}={}&-i[\hat H,\hat\rho]
 +\kappa_{\rm c}(\bar n_{\rm th}+1)\,\mathcal L_{\hat a}\hat\rho
 +\kappa_{\rm c}\bar n_{\rm th}\,\mathcal L_{\hat a^\dagger}\hat\rho \nonumber \\
&+\sum_{j=1}^{N}\left[
 \gamma\,\mathcal L_{\hat s_j^-}
 +\gamma_\phi\,\mathcal L_{\hat s_j^z}
 +(W+\gamma)\,\mathcal L_{\hat s_j^+}
 \right]\hat\rho,
\end{align}
where the coherent dynamics is governed by the Hamiltonian given in equation~\eqref{Hamiltonian} and the dissipation processes are described by the Lindblad superoperators $\mathcal L_{\hat o}\hat\rho =\hat o\hat\rho\hat o^\dagger -\left\{\hat o^\dagger\hat o,\hat\rho\right\}/2$, with 
 $\kappa_{\rm c}$ being the cavity decay rate, $\bar n_{\rm th}$ being the thermal photon occupation, $\gamma$ being the spin relaxation rate, $\gamma_\phi$ being the spin dephasing rate, and $W$ being the incoherent pump rate. Note that the thermally induced spin-pumping term under ambient conditions is included.
%The Hamiltonian of the system reads (with $\hbar$ set as unity)
%\begin{equation}
%\hat{H}=\omega_{\rm c}\hat{a}^\dagger\hat{a} + \sum_{j=1}^{N}\omega_j\hat{s}_j^z  +ig\left(\hat{a}^\dagger\hat{S}_{\rm tot}^- -\hat{a}\hat{S}_{\rm tot}^+\right),
%\end{equation}
%where $\hat{a}^{\dagger}$ ($\hat{a}$) is the creation (annihilation) operator of the quantized cavity field, and $\hat{s}_{j}^{z}$ is the $z$-component of the $j$-th spin.
%We also introduce collective spin operator $\hat{S}_{\rm tot}^\pm=\sum_{j=1}^N\hat{s}_{j}^\pm$ with $\hat{s}_{j}^\pm$ the raiser/lower operators of the $j$-th spin.
%The the single-spin cavity coupling strength is $g$, which gives the collective coupling strength $G\equiv\sqrt{N}g$.
%The quantities $\omega_j$  and $\omega_{\rm c}$  denote the $j$-th spin-transition and cavity-resonance frequencies, respectively. 

The solution of the master equation for $\sim 10^{14}$ spin-1/2's and a cavity mode is formidable when the spins have different frequencies even in the mean-field approximation. For numerical calculations, we divide the inhomogeneously broadened spin ensemble into $2M+1$ sub-ensemble, each of which has a width of frequency distribution narrow enough for all $N_{\mu}$ spins in the sub-ensemble to be treated to have a common frequency $\omega_{\mu}$ (for $\mu=M,M-1,\ldots,-M$). {The fraction of spins in the $\mu$-th sub-ensemble $p_{\mu}=N_{\mu}/N$  is determined by a Gaussian distribution centered at $\omega_{\rm s}$ with a standard deviation $\sigma/\sqrt{2}$}. Thus, the master equation is reduced to one for $2M+1$ collective spins ($S_{\mu}\le N_{\mu}/2$) and one cavity mode. We increase the number of sub-ensembles $2M+1$ until the solution converges.

We solved the coupled master equations in the mean-field approximation. 
In a frame rotating at the cavity frequency $\omega_{\rm c}$, the resulting mean-field equations are
\begin{align}
\frac{da}{dt}
 &= -\frac{\kappa_{\rm c}}{2}a +G\sum_{\mu=-M}^{M}p_{\mu}S_{\mu}^-+\sqrt{\frac{\kappa_{\rm c}\bar{n}_{\rm th}}{2N}}\,\xi(t),\label{eq:mf_cavity}\\
\frac{dS_{\mu}^-}{dt}
 &= -\left(\frac{\kappa_{\rm s}}{2}+i\omega_{\mu}-i\omega_{\rm c}\right)S_{\mu}^-+2GS_{\mu}^za,\label{eq:mf_transverse}\\
\frac{dS_{\mu}^z}{dt}
 &= \frac{W}{2}-(W+2\gamma)S_{\mu}^z -G \left(a^*S_{\mu}^-+aS_{\mu}^+\right),
\end{align}
where $
S_{\mu}^{\pm,z}\equiv \sum_{j\in \mu}\langle\hat s_j^{\pm,z}\rangle/{N_{\mu}}$ is the normalized  coherence/polarization of the $\mu$-th spin sub-ensemble
(with $\sum_{j\in \mu}$ denoting the the summation over the sub-ensemble), $S_{\mu}^+=(S_{\mu}^-)^*$, $a\equiv {\langle\hat a\rangle}/{\sqrt N}$ is the normalized photon amplitude, and $\kappa_{\rm s}=W+2\gamma+\gamma_\phi$ is the total dissipation rate of the collective spin of each sub-ensemble. 
 %These normalizations keep both the cavity-field amplitude and the spin polarizations of order unity to improve the numerical conditioning.
We account for the unavoidable thermal fluctuations of the cavity field in Eq.~(\ref{eq:mf_cavity}) with a stochastic Langevin term. Here $\bar{n}_{\rm th}$ denotes the mean thermal photon occupation of the cavity mode, $\xi(t)=\xi_x(t)+i\xi_y(t)$ is a complex Gaussian white noise satisfying $\langle\xi_a(t)\rangle=0,\langle\xi_a(t)\xi_b(t')\rangle=\delta_{ab}\delta(t-t'),$ with $a,b\in\{x,y\}$. 
To reduce computational cost, we neglect the small stochastic Langevin terms for spin dynamics. The maser power is computed as $P = N\hbar \omega_{\rm c}\vert a\vert^2\kappa_{\rm c}$.
Further details of the simulations are provided in the Supplementary Information~\ref{SI1} and \ref{SI5}.  

We compare the experimental and simulated results in Figs.~\ref{Figure2},~\ref{fig_switchoff} and~\ref{fig:phase_diagram},  and Extended Data Figs.~\ref{fig_interval_amp_corr} and \ref{fig_s7}. 
In all cases, the simulations reproduce the main features of the experimental data with reasonable quantitative agreement.
The simulation parameters are set to experimentally measured values, except for the incoherent pump rate, $W$. To avoid uncertainties arising from the complex scattering, reflection and absorption of pump laser within the sample, we calibrate the experimental and simulated pump rates using the normalized ratio $W/W_{\rm th}$, where the masing threshold at $Q=40,000$ gives $W_{\rm th}$=33~Hz.

%Our model captures the essential physics underlying the observed superradiant phases, which arise from the emergence and breakdown of synchronization and macroscopic quantum coherence across the spin ensemble. The remaining discrepancies may partly originate from the idealized Gaussian spin-frequency distribution used in the simulations, which does not fully capture the inhomogeneous broadening of the experimental sample. 
% Spin–spin interactions within the dense ensemble may generate nontrivial spin–spin and spin–photon correlations that warrant further investigation.

%%=============================================%%
%% Extended data Figures %%
%%=============================================%%
{\centering\subsection*{Extended Data Figures}\par}
\renewcommand{\figurename}{Extended Data Fig.}
\renewcommand{\thefigure}{E\arabic{figure}}
\setcounter{figure}{0} 

\begin{figure}[H]
\centering
\includegraphics[width=0.95\linewidth]{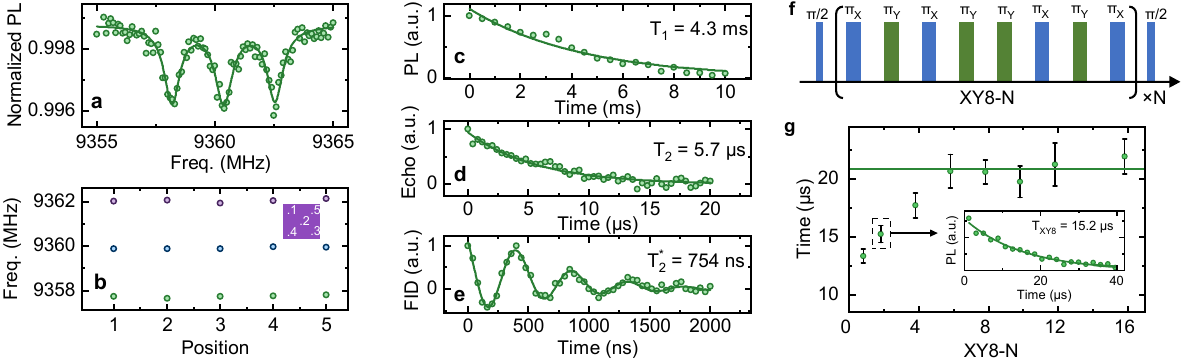}
\caption{\label{fig_s1} \textbf{Characterization of NV centre spins}.
\textbf{a,} Optically detected magnetic resonance (ODMR) spectrum of the NV$^-$  centres in the diamond, measured in situ in the maser setup at low microwave power. The three hyperfine peaks arise from the coupling of the NV electron spin to the  $^{14}\mathrm{N}$ nuclear spin. Lorentzian fits to these peaks yield an average linewidth of $0.8~\mathrm{MHz}$.
\textbf{b,} ODMR frequencies of the three hyperfine resonances at five positions across the diamond  with size $\sim 2 \times 2~\rm{mm^2}$ (inset), showing a standard deviation of $0.065~\mathrm{MHz}$ and a maximum spread of $0.18~\mathrm{MHz}$ (corresponding to a variation of magnetic field about $0.064~\mathrm{Gs}$).
\textbf{c,} Longitudinal relaxation of the spins, with a relaxation time $T_1$ = 4.3 ms. 
\textbf{d,} Hahn echo signal of the spins, with a dephasing time $T_2 = 5.7$~\textmu s.
\textbf{e,} Free-induction decay of the spins, with a decay time $T_{2}^{*} = 754$~ns, which indicates an inhomogeneous broadening of 0.42 MHz.
\textbf{f,} XY8-$N$ dynamical decoupling pulse sequence. \textbf{g,} Spin dephasing time as a function of the number of pulses in the dynamical decoupling sequence in \textbf{f}, saturating at approximately $21$~\textmu s, corresponding to an [NV$^{-}$] concentration of $2.7\pm0.1$~ppm~\cite{zhang2026unraveling}. Inset, spin dephasing under XY8-2, with the dephasing time extracted as $15.2$~\textmu s.}
\end{figure}

\begin{figure}[H]
\centering
\includegraphics[width=0.95\linewidth]{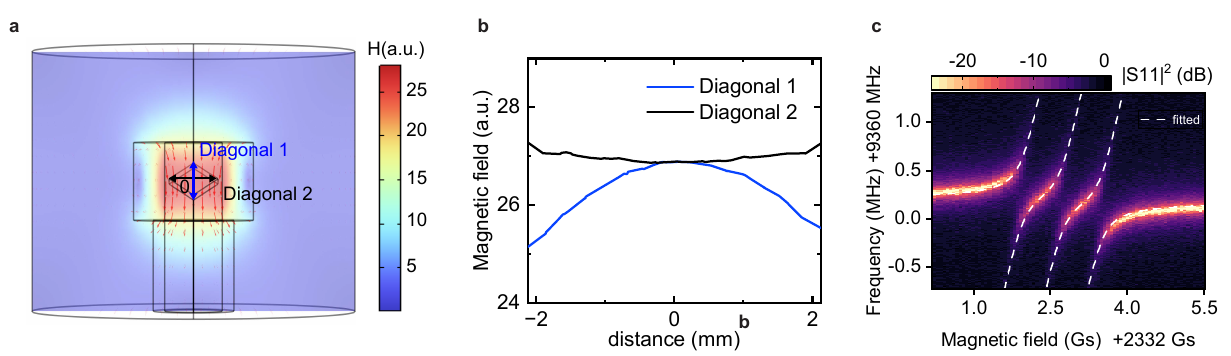}
\caption{\label{fig_s2} 
\textbf{Spin-cavity coupling.} \textbf{a,} Magnetic field distribution of the cavity mode obtained by simulation using COMSOL. The contour color represents the magnetic field strength, and the red arrows indicate the field direction. \textbf{b,} The magnetic field magnitude of the cavity mode along the two diagonals of the diamond indicated in \textbf{a}, showing a variation of less than $8\%$ across the sample.
\textbf{c,} Spectrum of cavity reflection signal ${\left|S_{11}\right|^2}$ as a function of the external magnetic field, measured using a Vector Network Analyzer (VNA) by tuning the transition $\lvert 0\rangle\leftrightarrow\lvert +1\rangle$
 of the NV$^{-}$ centres through the cavity resonance, with a pump power of 1773~mW (which nearly saturates the spin polarization) and a loaded $Q$ factor of 21{,}000. The dashed lines are fitting with a collective spin-cavity coupling $G/2\pi=0.32$~MHz.
}
\end{figure}

\begin{figure}[H]
\centering
\includegraphics[width=0.95\linewidth]{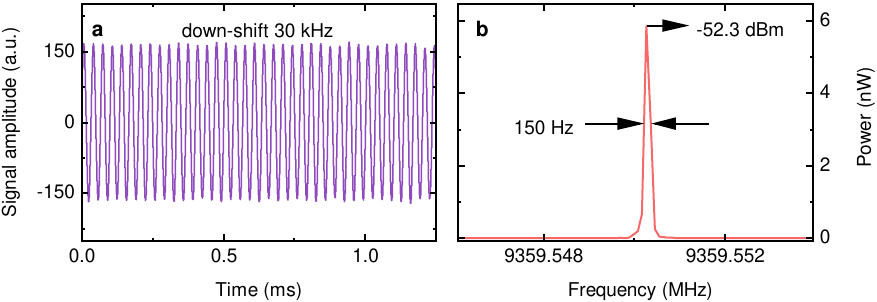}
\caption{\label{fig_maxPower} 
\textbf{Maser with maximum output power.} The maximum maser power output is measured at a cavity quality factor Q=22,500, a laser pump power of 1.7 W, a spot
diameter of 1.5 mm {(corresponding to a pump rate $W/W_{\rm th}\sim 200$)}, and a spin-cavity detuning $\Delta/2\pi=-0.52 ~\rm{MHz}$. \textbf{a} and \textbf{b} show the time-domain signal and the spectrum, respectively.  The maximum output power $P_{\rm out}^{\rm max}\simeq-52.3~\mathrm{dBm}$ yields an intracavity photon number $P_{\rm out}^{\max}/\hbar\omega_{\rm{m}}\kappa_{\rm{out}} \sim 1.45 \times 10^9$.
}
\end{figure}

\begin{figure}[H]
\centering
\includegraphics[width=0.95\linewidth]{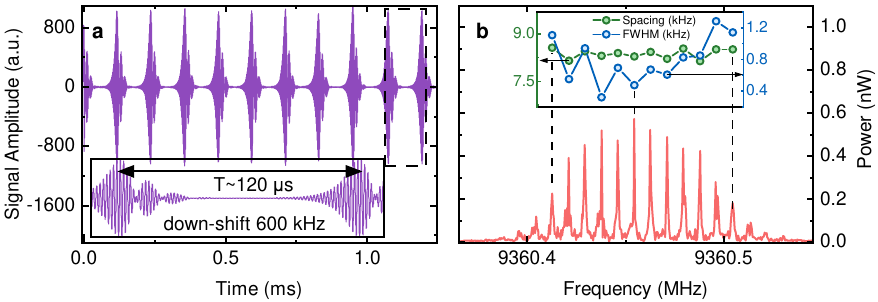}
\caption{\label{fig_partial_synchronization} 
{\textbf{Partial synchronization of spin coherence in the PAM regime.}} \textbf{(a)} and \textbf{(b)} show the time-domain signal and the spectrum of the maser in the PAM regime, respectively. The spin-cavity detuning $\Delta/2\pi=0 ~\rm{MHz}$ and other parameters are the same as Fig.~\ref{fig_maxPower}.
The coherent oscillation of the signal above the noise floor is observed throughout the $120$~\textmu s interval (see inset in \textbf{a}), which indicates persistent macroscopic coherence. {The frequency comb is well resolved with fineness (the ratio of comb separations to peak linewidths) being about 10 (shown in inset of \textbf{b}), which indicates phase memory over 10 periods (that is, 1.2~ms).} Note that the superradiance pulse in each period is further modulated with three sub-pulses. Such multi-peak structure means that the spin sub-ensembles are not fully synchronized. Instead, as in the present case, the spins are self-organized into individually synchronized groups. The interference between these groups gives rise to the multi-peak structure in each period. 
%{Note that the multi-peak structure within each superradiance pulse originates from the dynamical oscillations of individual spin sub-ensembles and mutual interference among these sub-ensembles. Nevertheless, the majority of spin sub-ensembles maintain synchronization, which is responsible for the formation of the observed frequency comb.}
}
\end{figure}

\begin{figure}[H]
\centering
\includegraphics[width=0.95\linewidth]{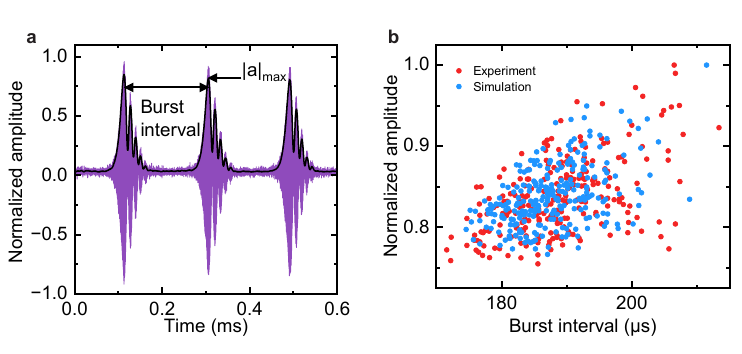}
\caption{\label{fig_interval_amp_corr} 
\textbf{Fluctuations of intervals and amplitudes of superradiant bursts.}
\textbf{a}, Measured signal (Purple line, I signal; Black line, smoothed amplitude) in the burst regime showing that the burst interval and peak amplitude $\left|a\right|_{\rm max}$ are random. \textbf{b}, Distributions and correlations of the burst intervals and normalized maximum amplitudes of 29 successive bursts measured in 9 experimental runs. In each run of experiments, the laser is turned off and then on to re-initialize the system and regenerate the burst sequence in a steady state. The simulation results are extracted from 266 successive bursts. The experiment conditions are the same as for Fig.~\ref{fig_switchoff}c, and the  parameters for the numerical simulation are the same as for Fig.~\ref{fig_switchoff}d. %\mynote{what are the parameters? They should be specified for all figures}
}
\end{figure}

\begin{figure}[H]    
\centering
\includegraphics[width=0.95\linewidth]{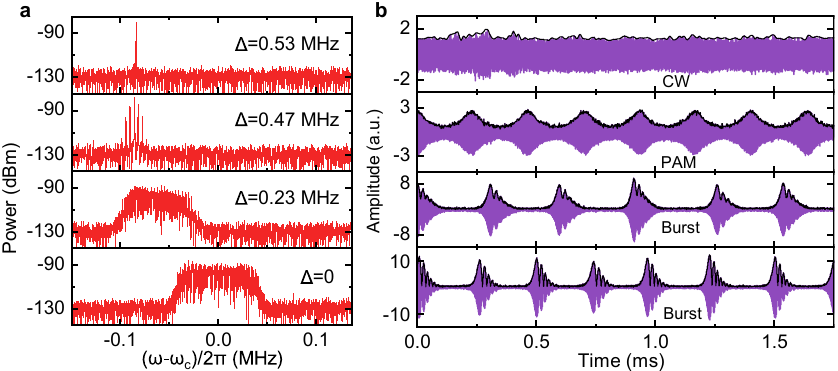}
\caption{\label{fig_s9} 
\textbf{Dependence of the superradiant maser on spin–cavity detuning.} \textbf{a} and \textbf{b}, Measured spectra- and time-domain signals for spin-cavity detuning $\Delta/2\pi=0.53$, $0.47$, $0.23$ and $0$~MHz, from top to bottom.  The pump rate is fixed at $W/W_{\rm th}=4.5$ and the cavity quality factor $Q=40,000$. The CW, PAM, and burst phases occur subsequently when the conditions go further above the threshold. In the burst phase, when the the spin-cavity detuning is non-zero, the interference between the signals from different groups of synchronized spins is not complete, as can be seen from the envelope (black line) in the 3rd row of \textbf{(b)}.
}
\end{figure}

\begin{figure}[H]
\centering
\includegraphics[width=0.95\textwidth]{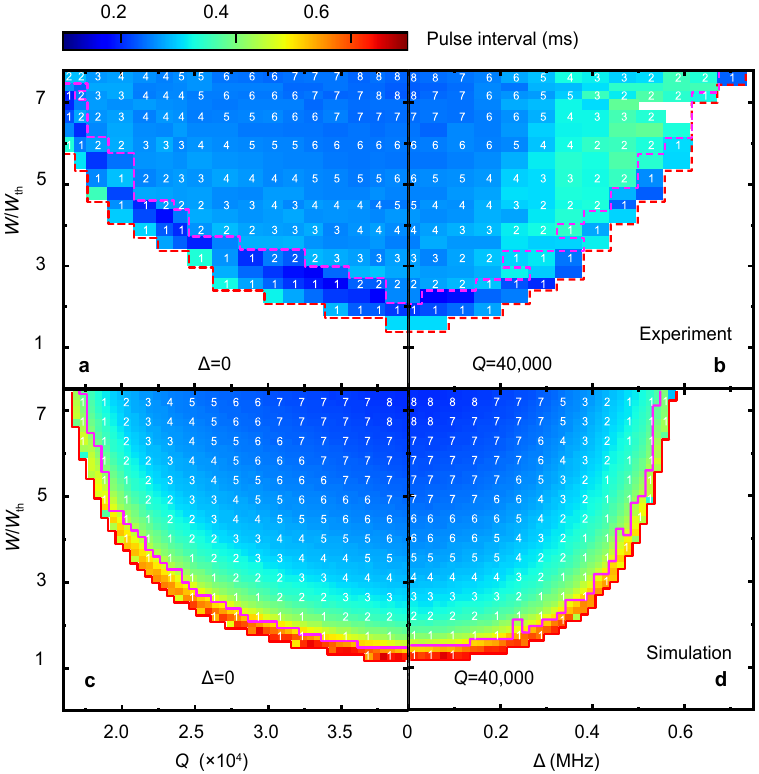}
\caption{\label{fig_s7}
\textbf{Modulation period or average interval between bursts in the PAM and burst phases.} \textbf{a} and \textbf{c}, Experimental and simulated results as functions of the $Q$ factor and the pump rate {(normalized by the threshold pump rate $W_{\rm th}$ at  $Q=40,000$ and $\Delta=0$)}, with the spin-cavity detuning fixed at $\Delta=0$. \textbf{b} and \textbf{d}, Same as \textbf{a} and \textbf{c} but as functions of the spin-cavity detuning and the pump rate, with the $Q$ factor fixed at $Q=40,000$. 
The numbers of the sub-peaks within each period/burst are marked.
Both the experimental and simulated results showed that the pulse interval decreased with increasing pump power, larger $Q$ factor, and smaller detuning. All other parameters are the same as in Fig.~\ref{fig:phase_diagram}.
}
\end{figure}



\newpage
{\centering\section*{SUPPLEMENTARY INFORMATION}\par}
\renewcommand{\figurename}{Fig.}
\renewcommand{\thefigure}{S\arabic{figure}}  
\renewcommand{\thesection}{Note~\arabic{section}}  
\renewcommand{\theequation}{S\arabic{equation}}  
\renewcommand{\thetable}{S\arabic{table}}  

\renewcommand{\theHfigure}{supp.figure.\arabic{figure}}
\renewcommand{\theHsection}{supp.section.\arabic{section}}
\renewcommand{\theHequation}{supp.equation.\arabic{equation}}
\renewcommand{\theHtable}{supp.table.\arabic{table}}

\setcounter{section}{0}
\setcounter{section}{0}
\setcounter{figure}{0}
\setcounter{equation}{0}
\setcounter{table}{0}

\section{Spin sub-ensembles and parameters}\label{SI1}

For the numerical calculations, the inhomogeneously broadened spin ensemble $\{\omega_j\}$ is divided into $2M+1$ sub-ensembles. All spins in the $\mu$-th sub-ensemble share the same transition frequency $\omega_{\mu}$. 
The sub-ensemble contains $N_{\mu}$ spins and has a population fraction $p_{\mu}=N_{\mu}/N$, with $\sum_{\mu=-M}^{M}p_{\mu}=1$. 
The frequencies $\omega_{\mu}$ are uniformly spaced within
\begin{equation}
\omega_{\mu}=\omega_{\rm s}+{\mu}\delta \in
[\omega_{\rm s}-\nu_{\rm cutoff},\,\omega_{\rm s}+\nu_{\rm cutoff}],
\end{equation}
for $\mu=-M,-M+1,\ldots,M$, 
where the frequency interval $\delta\equiv \nu_{\rm cutoff}/{M} $ and the cutoff is chosen to be $\nu_{\rm cutoff}=3{\sigma}/{\sqrt{2}}$ in our simulation.
Experimentally, the spin-cavity detuning is tuned over $\Delta/2\pi\equiv (\omega_{\rm s}-\omega_{\rm c})/2\pi\in[-0.8,\,0.8]~{\rm MHz}$ by sweeping the external magnetic field.
To represent the inhomogeneous distribution, the number of spins in each sub-ensemble is
\begin{equation}
N_{\mu}= N\frac{\delta}{\sigma\sqrt{2\pi}} e^{-\left(\omega_{\mu}-\omega_{\rm s}\right)^2/2\sigma^2}.
%=\frac{F(\omega_{\mu+1/2})-F(\omega_{\mu-1/2})}{F(\omega_{\rm s}+3\frac{\sigma}{\sqrt{2}})-F(\omega_{\rm s}-3\frac{\sigma}{\sqrt{2}})}, F(\omega)=\frac{1}{2}\left[1+\operatorname{erf}\left(\frac{\omega-\omega_{\rm s}}{\sigma}\right)\right],
\end{equation}
%where $\omega_{\mu\pm1/2}$ are the bin boundaries.
%We calibrate $\sigma$ by requiring the discretized free-induction signal to reproduce the Gaussian decay envelope, $\sum_{\mu} p_{\mu} e^{-i(\omega_{\mu}-\omega_{\rm s})t}\simeq\exp[-(t/T_2^*)^2]$, which gives the inhomogeneous linewidth $\sigma=2/T_2^*$.
Unless otherwise stated, we use $M=160$, corresponding to $321$ sub-ensembles, which is sufficient to ensure
numerical convergence of the mean-field dynamics.

All fixed model parameters derived from experimental measurements are listed in Supplementary Table~\ref{table_parameters}. The pump rate $W$, the spin-cavity detuning $\Delta \equiv \omega_{\rm s} - \omega_{\rm{c}}$, and the cavity quality factor $Q$ are experimentally tunable and are specified for each simulation. 
\begin{table}[h]
\centering
\caption{Parameters used in the numerical simulations. Fixed parameters are derived from experimental measurements.}
\renewcommand{\arraystretch}{1.3}
\setlength{\tabcolsep}{10pt}
\begin{tabular}{ll}
\hline
Parameters & Description \\
\hline
$\omega_{\rm c}/2\pi=9.360{\,\rm GHz}$ & Cavity resonance frequency \\
$\kappa_{\rm c}=\omega_{\rm c}/Q$ & Cavity leakage rate, calculated from the cavity quality factor $Q$ \\
$\bar{n}_{\rm th}=670$ & Mean thermal photon occupation of the cavity mode at $T=300~{\rm K}$ \\
$N=1.40\times10^{14}$ & Effective number of spins coupled to the cavity mode\\
$\gamma/2\pi=37.0{\,\rm Hz}$ & Spin longitudinal relaxation rate corresponding to $T_1=4.3 {\,\rm ms}$ \\
$\gamma_\phi/2\pi=55.8{\,\rm kHz}$ & Spin pure dephasing rate corresponding to $T_2=5.7{\,\rm \mu s}$ \\
$\sigma/2\pi=0.422$~MHz & Inhomogeneous linewidth estimated from $T_2^*=0.75{\,\rm \mu s}$ \\  
$G/2\pi=0.32$~MHz & Collective spin-cavity coupling strength \\
\hline
\end{tabular}
\label{table_parameters}
\end{table}

\section{Superradiance}\label{SI2} 

For a homogeneous spin ensemble comprising a single sub-ensemble ($M=0$), the system enters the superradiant masing phase when the effective  cooperativity is above the threshold~\cite{xiao2026squeezed},
\begin{equation}
C\equiv\frac{4G^2}{\kappa_{\rm c}\kappa_{\rm s}}\frac{W}{W+2\gamma}\frac{\Gamma^2}{4\Delta^2+\Gamma^2}>1,
\label{eq:effective_cooperativity}
\end{equation}
where $\kappa_{\rm s}=W+2\gamma+\gamma_\phi+\sigma$ is the total spin dissipation rate including the spin dephasing induced by the inhomogeneous broadening, and $\Gamma=\kappa_{\rm c}+\kappa_{\rm s}$ is the total dissipation rate. For a homogeneous spin ensemble resonant with the cavity at $Q=40,000$, the pump threshold is calculated to be $W_{\rm th}/2\pi=27.8{\,\rm Hz}$. 
For an inhomogeneously broadened ensemble with $M\gg1$, the system undergoes the same transition, but the distribution of spin transition frequencies modifies the effective cooperativity and consequently shifts the masing threshold, which has no closed-form expression and is determined numerically as $W_{\rm th}/2\pi=33.0{\,\rm Hz}$.

\begin{figure}[H]
\centering
\includegraphics[width=0.95\textwidth]{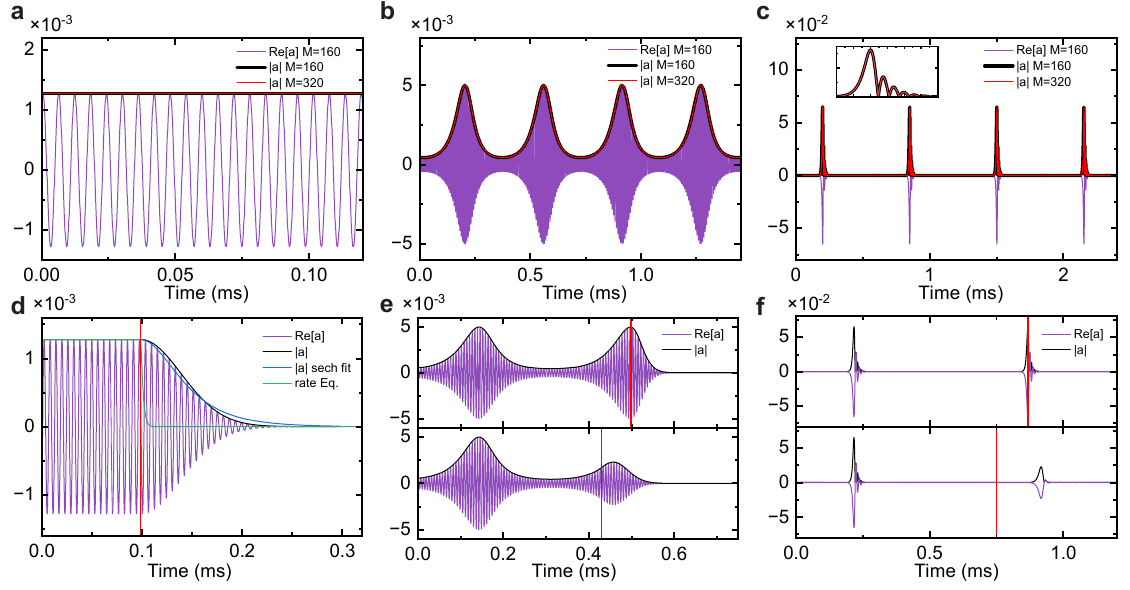}
\caption{\label{fig_s3} 
\textbf{Numerically calculated cavity field under continuous pumping and after pump-off.}  
The cavity quality factor  $Q=26,300$  and the pump rate $W/W_{\rm th}=6.45$. 
The cavity field under continuous pump in the CW maser regime (\textbf{a}, with $\Delta/2\pi=0.470$~MHz), the PAM regime (\textbf{b}, with $\Delta/2\pi=0.434$~MHz), and the superradiant-burst regime (\textbf{c}, with $\Delta=0$). The purple curves show the real part of the cavity field $\operatorname{Re}[a]$ , while the black and red curves show its amplitude $\lvert a\rvert$ , calculated with the spin ensemble divided into 321 or 641 spin sub-ensembles ($M=160$  or $320$, respectively). All other parameters are the same as those in Tab.~\ref{table_parameters}. The thermal photon number is set as $\bar{n}_{\rm th}=0$.
(\textbf{d}/\textbf{e}/\textbf{f}) shows the cavity field after the pump is switched off (at times marked by the vertical red lines), calculated with the same parameters as in (\textbf{a}/\textbf{b}/\textbf{c}) and $M=160$. The blue curves are fits to the superradiant pulse profile $\lvert a(t)\rvert\propto\operatorname{sech}[(t-t_0)/\tau_{\mathrm{SR}}]$, and the green curves show the emission dynamics predicted by the rate-equation model. 
}
\end{figure}

% Re-check the latter used parameters in the simulation
At a fixed cavity quality factor $Q=26,300$ and a pump rate $W/W_{\rm th}=6.45$, we show three characteristic phases of the maser for different spin-cavity detunings (Fig.~\ref{fig_s3}a-c), namely, the CW phase for $\Delta/2\pi=0.470$~MHz, the PAM phase for $\Delta/2\pi=0.434$~MHz, and the burst phase for $\Delta=0$. 
 The cavity-field amplitudes $|a|$ obtained for $M=160$ (black solid lines) and $M=320$ (red solid lines) overlap almost perfectly except for a constant temporal offset, which confirms the convergence of the mean-field dynamics with respect to the discretization of the spin ensemble. 
The calculation for fixed spin-cavity detuning and varying pump rate $W/W_{\rm th}$ yield similar results.

To demonstrate the superradiance nature of the maser, we perform a series of measurements after the pump is turned off and compared the data with predictions from both the superradiant maser model and the rate equations. The rate equations for the dynamics of independent spins are~\cite{carmichael2013statistical}
\begin{subequations}
\begin{align}
    &\frac{dP_{e}}{dt}=w(1-P_e)-\gamma P_{e}-\Gamma_{\rm bare}\left(n+1\right)P_{e}+\Gamma_{\rm bare}n\left(1-P_{e}\right),\\
    &\frac{dn}{dt}=N\Gamma_{\rm bare}\left[\left(n+1\right)P_{e}-n\left(1-P_{e}\right)\right]-\kappa_{\rm c} n,
    \label{RateEquation}
\end{align}
\end{subequations}
where $P_e$ is the excited-state population of a single spin, $n$ is the mean intra-cavity photon number, and $\Gamma_{\rm bare}=4g^2/\kappa_{\rm s}$ is the the cavity-mediated transition rate of a single spin. In the CW regime, when the pump is switched off, the characteristic superradiant emission can be roughly fitted by $|a(t)|\propto\operatorname{sech}[(t-t_0)/\tau_{\rm SR}]$, as shown in Fig.~\ref{fig_s3}d. The rate equations produces a much faster decay with the same set of parameters. In the PAM regime, when the pump is switched off near a minimum, the signal first rises and then decreases, resembling a revival of the previous period but with reduced amplitude. This phenomenon cannot be explained by simple rate equations, since, after the pump is switched off, the emission amplitude calculated by the rate equations monotonically decreases. In the superradiant maser model, the revival behavior is well understood since the macroscopic collective spin coherence is generated and transferred back to the cavity field. Similarly, revivals are also observed in the superradiant burst regime. These results resemble the experimental observation, and confirm that the maser signal is superradiant and that macroscopic coherence is established in the spin ensemble.

\section{Effects of thermal noise}\label{SI3}

\begin{figure}[H]
\centering
\includegraphics[width=0.95\textwidth]{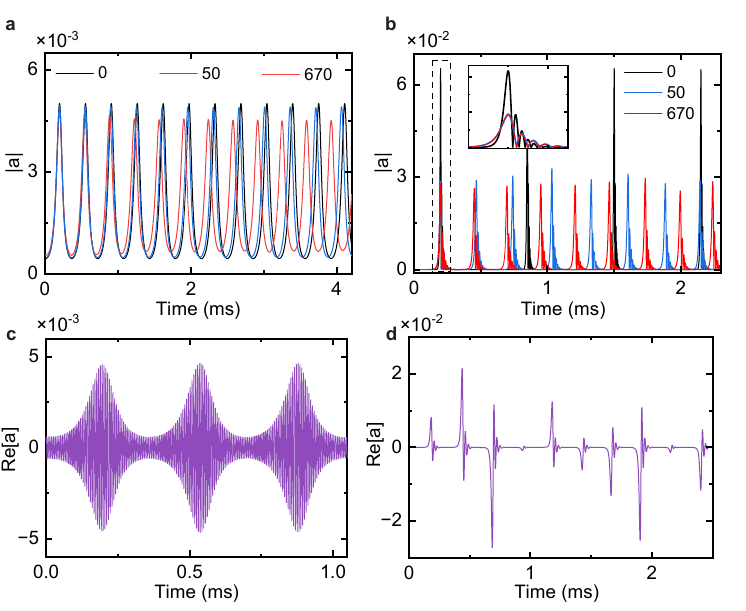}
\caption{\label{fig_s5}
\textbf{Effects of thermal noise on the cavity-field dynamics.} The cavity-field amplitude $|a|$ in the PAM (\textbf{a}) and superradiant burst (\textbf{b}) regimes for thermal photon occupation $\bar{n}_{\rm th}=0$  (black), $50$  (blue), and $670$  (red). The inset in \textbf{b} enlarges the region bounded by the vertical dashed lines. 
The cavity field ${\rm Re[}a{\rm ]}$  shows coherent oscillation in the PAM regime (\textbf{c}), and  exhibits random phase jumps between successive bursts in the superradiant burst regime (\textbf{d}). In \textbf{(c)} and \textbf{(d)}, the thermal photon population is $\bar{n}_{\rm th}=670$.
All other parameters are the same as those in Fig.~\ref{fig_s3}.
}
\end{figure}

To check the effects of thermal noises, we compare the cavity-field dynamics for different thermal photon occupations, $\bar{n}_{\rm th}=0$, $50$ , and $670$. In the PAM regime, the macroscopic coherence 
persists during the spin polarization buildup intervals, the thermal noise only weakly perturbs the modulation period and the amplitude (Fig.~\ref{fig_s5}a), and the phase of the maser is robust (Fig.~\ref{fig_s5}c). Such robustness of phase coherence indicates a time-crystalline order that is resilient to thermal fluctuations. By contrast, in the burst regime, 
the spin coherence vanishes during the polarization buildup intervals and the superradiance needs to be triggered spontaneously in each cycle, which is sensitive to noises as the thermal noises can seed the superradiance bursts. Therefore, the burst intervals and amplitudes strongly depend on the thermal noises (or thermal photon population) and the relative phases between subsequent bursts are totally random, as observed in the numerical simulation shown in  Fig.~\ref{fig_s5}b,d. It is worth noting that the observed randomness in intervals, amplitudes, and relative phases of the bursts can also result from high pump rates even in absence of thermal photons.

\section{Synchronization of macroscopic coherence}\label{SI4} 
\begin{figure}[H]
\centering
\includegraphics[width=0.95\textwidth]{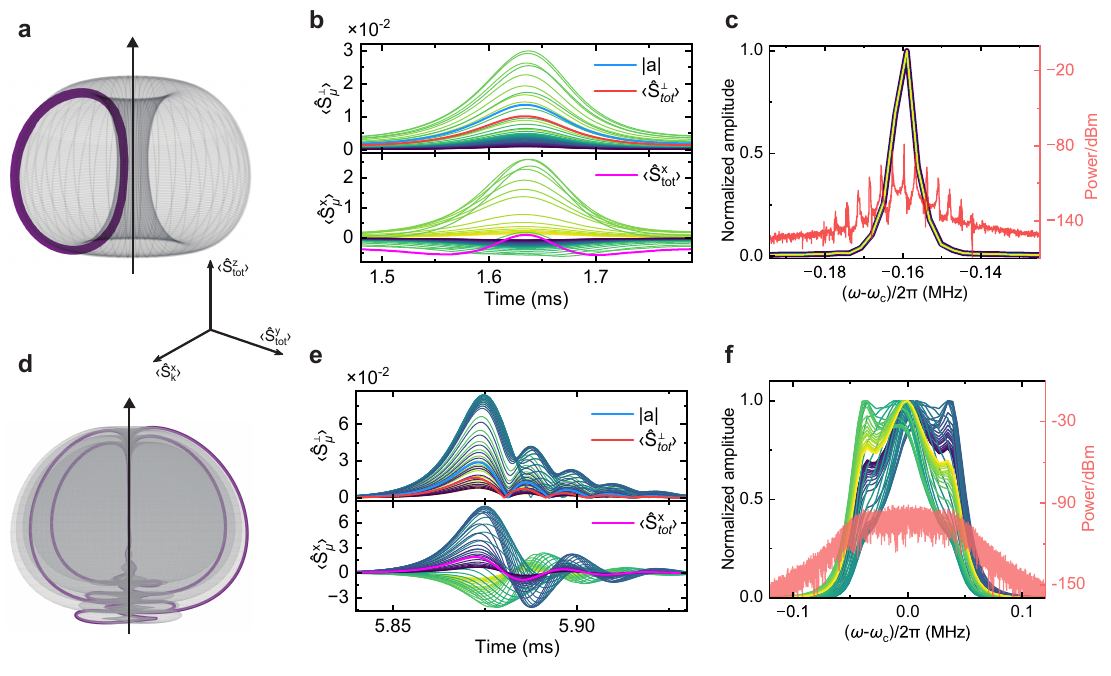}
\caption{\label{fig_s4}
\textbf{Collective spin dynamics with and without synchronization in different phases of the superradiant maser.} (\textbf{a}/\textbf{b}/\textbf{c}) and (\textbf{d}/\textbf{e}/\textbf{f}) show the spin dynamics and spectrum calculated for parameters in the PAM and burst regimes, respectively.
\textbf{a} and \textbf{d}, Trajectories of the total collective-spin Bloch vector $( \langle \hat{S}_{\rm tot}^{x}\rangle,\langle \hat{S}_{\rm tot}^{y}\rangle,\langle \hat{S}_{\rm tot}^{z}\rangle)$, over several pulse intervals.
\textbf{b} and \textbf{e}, Time evolution of the transverse spin magnitude $\langle \hat{S}_{\mu}^{\perp}\rangle$ and the $x$-component $\langle \hat{S}_{\mu}^{x}\rangle$ for 55 representative spin sub-ensembles during a single period/burst. The gradient-colored curves distinguish the different sub-ensembles. For comparison, the cavity-field amplitude $|a|$, the total transverse spin magnitude $\langle \hat{S}_{\rm tot}^{\perp}\rangle$, and the total $x$-component $\langle \hat{S}_{\rm tot}^{x}\rangle$ are shown in blue, red, and magenta curves, correspondingly.
\textbf{c} and \textbf{f}, Normalized spin spectra of the representative spin sub-ensembles evaluated over a single pulse. The gradient color coding is identical to that in (\textbf{b}/\textbf{e}), such that the same color denotes the same sub-ensemble. The cavity-field power spectrum evaluated over the full pulse sequence are shown by the solid red curves (in units of dBm). All parameters are the same as those in Fig.~\ref{fig_s3}.
}
\end{figure}

The three maser phases differ in the degree of synchronization and macroscopic coherence across the spin ensemble. In the CW  and PAM phases, the total collective-spin Bloch vector $\left( \langle \hat{S}_{\rm tot}^{x}\rangle, \langle \hat{S}_{\rm tot}^{y}\rangle, \langle \hat{S}_{\rm tot}^{z}\rangle\right)$ evolves almost deterministically (Fig.~\ref{fig_s4}a), which indicates robust macroscopic coherence and synchronization of all of the spins in the ensemble. The transverse spin magnitude $\langle \hat{S}_{\mu}^{\perp}\rangle$ and the $x$ component $\langle \hat{S}_{\mu}^{x}\rangle$, evaluated over a single pulse interval, further show that spin sub-ensembles with different transition frequencies become frequency-locked with fixed relative phases in a common rotating reference frame $\omega_{\rm r}=-0.16$~MHz  (Fig.~\ref{fig_s4}b). In contrast, in the burst phase, synchronization across the whole spin ensemble is not maintained. Within a burst, synchronization and coherence are gradually lost and eventually vanish at the end. The distinct dynamics of $\langle\hat S_{\mu}^\perp\rangle$  and $\langle\hat S_{\mu}^x\rangle$  among spin sub-ensembles leads to the interference effects that shape the overall signal (Fig.~\ref{fig_s4}f). {After the burst, the transverse spin polarization, i.e., the macroscopic coherence, is lost.} The spin polarization needs to be re-built up by incoherent pump. The next burst follows different and spontaneously selected trajectories (Fig.~\ref{fig_s4}d).

\section{Numerical method}\label{SI5} The mean-field equations are integrated using an adaptive Runge-Kutta method (RK45), with relative and absolute tolerances of $10^{-7}$ and $10^{-10}$, respectively. Each simulation spans $0.5~{\rm s}$, substantially longer than all relevant intrinsic relaxation timescales of the model. To reduce the computational cost, the stochastic Langevin term is activated at $t=0.46~{\rm s}$, after the deterministic dynamics have approached their long-time attractor. The system is subsequently evolved with thermal noise for $40~{\rm ms}$. The first $20~{\rm ms}$ of the stochastic evolution is discarded as a noise-equilibration transient, and the final $20~{\rm ms}$ is used for data analysis. Within this analysis window, the solution is evaluated on a uniform temporal grid with a spacing of $0.1~\mu{\rm s}$, corresponding to a sampling rate of $10~{\rm MHz}$, sufficient to resolve the fastest relevant dynamics. For noisy simulation, we set $\bar{n}_{\rm th}=670$ to account for the room-temperature thermal noise associated with the microwave cavity. For each set of parameters, the stochastic dynamics are obtained from a single realization of $\xi(t)$, without ensemble averaging over noise realizations.

%\end{appendices}

% \bibliographystyle{sn-basic}   % Nature官方bst样式，自动上标
\bibliography{sn-bibliography}% common bib file

\end{document}